\documentclass[fleqn,usenatbib,useAMS]{mnras}

\usepackage[T1]{fontenc}
\usepackage{ae,aecompl}

\usepackage{graphicx}	
\usepackage{amsmath}	
\usepackage{amssymb}	

\newcommand{\bz}{\left< B_{\rm z} \right>}

\title[Sharp-lined Ap stars]{Magnetic field measurements of sharp-lined Ap stars
}

\author[ J\"arvinen et al.\ 2022]{
S.~P.~J\"arvinen$^{1}$\thanks{E-mail: sjarvinen@aip.de},
S.~Hubrig$^{1}$,
R.~Jayaraman$^{2}$,
I.~Ilyin$^{1}$,
and M.~Sch\"oller$^{3}$
  \\
$^{1}$Leibniz-Institut f\"ur Astrophysik Potsdam (AIP), An der Sternwarte~16, 14482~Potsdam, Germany\\
$^{2}$MIT Kavli Institute and Department of Physics, 77 Massachusetts Avenue, Cambridge, MA 02139, USA\\
$^{3}$European Southern Observatory, Karl-Schwarzschild-Str. 2, 85748 Garching, Germany\\
}
\date{Accepted 2022 August 19. Received 2022 August 18; in original form 2022 July 7 }

\pubyear{2022}

\begin{document}
\label{firstpage}
\pagerange{\pageref{firstpage}--\pageref{lastpage}}
\maketitle

\begin{abstract}
Previous observations suggested that Ap and Bp stars exhibit a bimodal
distribution of surface magnetic field strengths and that actually only few 
or no stars exist with magnetic dipole field strengths below 300\,G down to 
a few Gauss. As the number of Ap and Bp stars currently known to possess weak
magnetic fields is not large, it is necessary to carry out additional
spectropolarimetric studies of Ap and Bp stars to prove whether the
assumption of the existence of a critical value for the stability of
magnetic fields is realistic. In this study, we present high-resolution
HARPS\-pol magnetic field measurements for a sample of Ap stars with sharp
spectral lines with a view to characterize the strengths of their magnetic
fields. Out of the studied seven sharp-lined stars, two stars, HD\,174779
and HD\,203932, exhibit a rather weak longitudinal magnetic field with
$\left< B_{\rm z}\right>=-45\pm3$\,G and
$\left< B_{\rm z}\right>=21\pm4$\,G, respectively. Additionally, TESS
observations were used to test previous conclusions on the differentiation
of rotation periods of Ap and Bp stars. Apart from HD\,189832 and
HD\,203932, all other studied sharp-lined stars have long rotation periods.
Since an explanation for the slow rotation of Ap stars is currently missing,
additional studies of slowly rotating Ap and Bp stars are necessary to
improve our understanding of the formation and evolution of Ap and Bp stars.
\end{abstract} 




\section{Introduction}
\label{sec:intro}

Large-scale organised magnetic fields with strengths ranging from several 
tens of Gauss to several kG are present in 10 to 15\% of all stars of 
spectral types O to early F 
\citep[e.g.][]{Grunhut2017, Schoeller2017, book2021}.
Magnetic fields cover the whole stellar surface, and their geometry often 
resembles a single dipole whose axis is inclined with respect to the stellar 
rotation axis. Understanding how these stars acquired magnetic fields and 
why they are only present in a fraction of upper main-sequence stars, as 
well as how these fields affect stellar evolution, will have significant 
implications for a wide range of astrophysical areas, from galactic 
evolution to exoplanets. Studies of upper main-sequence Ap and Bp stars, 
constituting the most populous group of magnetic early-type stars, are also 
of great importance because such stars display the most extreme 
manifestations of the effects of magnetic fields on stellar atmospheres. 
Specifically, they exhibit strong abundance anomalies, including both the 
horizontal accumulation and vertical stratification of abundances of various 
chemical elements. 

Based on mean longitudinal magnetic field measurements (i.e.\ measurements of 
the line-intensity weighted average over the stellar disc of the component 
of the magnetic vector along the line of sight) of a small sample of 28 
magnetic Ap and Bp stars with poorly constrained magnetic field strengths,
\citet{Auriere}
concluded that there exists a critical dipole field strength, 
$B_{\rm d}\approx300$\,G, which corresponds to the minimum field strength 
for a star to maintain the stability of its magnetic field. Only two stars 
in the sample of 28 Ap stars showed a dipole strength below 300\,G. The 
authors suggested that the magnetic dichotomy of intermediate-mass stars -- 
i.e., a dichotomy in the distribution of the observed magnetic field between 
the kG dipoles of Ap and Bp stars and the sub-Gauss magnetism of Vega and 
Sirius -- arise from the development of non-axisymmetric instabilities 
separating stable strong field configurations observed in Ap and Bp stars
from unstable weaker field configurations, whose surface average field 
becomes very weak after the destabilization. However, in contrast to this 
study, a number of  Ap and Bp stars have been reported to possess very 
weak longitudinal magnetic fields on the order of a few tens of Gauss 
\citep[e.g.,][]{Donati1990, Donati2006, Alecian2016}.
Also, for higher mass Bp stars, rather weak magnetic fields were reported by 
\citet{Fossati2015}.
As discussed by 
\citet{JerCan},
such weak fields can be consistent with dynamo fields generated in 
subsurface convection zones. Since the reported dichotomy may be due to 
observational incompleteness, there is certainly a need for more 
representative studies of Ap and Bp stars. Such studies are especially 
important for the theoretical understanding of the origin of magnetic Ap and 
Bp stars.

The other important question for the understanding of the origin of magnetic 
Ap and Bp stars with sharp lines is whether the majority of stars with weak 
magnetic fields are slow rotators. If these stars are not observed close to 
the rotation pole, then we expect that a fraction of stars with very 
sharp spectral lines have longer rotation periods. The occurrence of very 
slow rotation in weakly magnetic Ap and Bp stars has not been investigated 
in detail yet, but is becoming an important subject of current studies of 
correlations between rotation rate and magnetic field strength, following 
the recent discovery of unbiased Transiting Exoplanet Survey Satellite
\citep[TESS;][]{TESSSPIE}
samples of slowly rotating Ap stars by 
\citet{MathysTESS, Mathys2022}.
According to 
\citet{Hubrig2007},
stronger magnetic fields tend to be found in hotter, younger and more 
massive stars, as well as in stars with shorter rotation periods.
\citet{Mathys2017} 
confirmed that Ap stars with very strong magnetic fields never achieve 
extremely slow rotation: Ap stars with $\langle B\rangle >7.5$\,kG have 
rotation periods shorter than 150\,d whereas Ap stars with 
$\langle B\rangle <7.5$\,kG have periods longer than 150\,d. Thus, the 
differentiation of rotation in Ap and Bp stars is a possible key to the 
understanding of the origin of their magnetic fields.

In this work, we discuss new mean longitudinal magnetic field measurements 
for a sample of sharp-lined Ap stars using the High Accuracy Radial velocity 
Planet Searcher polarimeter, 
\citep[HARPS\-pol;][]{Harps} 
fed by the ESO 3.6-m telescope. In the spectra of several of the Ap and Bp 
stars obtained in the framework of a systematic search for Ap stars with 
resolved magnetically split lines, spectral lines appear very sharp. This 
indicates that magnetic line broadening, which is proportional to the 
absolute value of the magnetic field strength, should be very small. For a 
number of stars in our sample, photometric data have recently been provided by 
the TESS mission, enabling us to search for variability in their light 
curves. As the surfaces of Ap and Bp stars are covered by long-lived 
chemical spots that rotate in and out of view, the light curves 
show rotational modulation. Usually, the light curves exhibit different 
shapes and amplitudes, depending on the size of the spots and their location 
with respect to the line of sight. The availability of space-based photometric 
observations is especially valuable in studies of Ap and Bp stars, as their 
rotation periods are most frequently determined from light curves.
 
Among the stars in our sample, magnetic field measurements are carried out 
for the first time for the sharp-lined stars HD\,89393, HD\,174779, and 
HD\,189832. In the spectra of HD\,138633, HD\,176196, HD\,203932, 
and HD\,217522, the spectral lines are also sharp and do not show any hint of 
resolution into their magnetic Zeeman components, but the possible presence of 
relatively weak magnetic fields was already discussed in previous studies.
The very slowly rotating star HD\,137949, which possesses a rather strong 
magnetic field and exhibits spectral lines resolved into magnetically split 
components, was previously reported to have a rotation period of more than 
14\,yr by 
\citet{Mathys2017}.
However, their published magnetic field measurements for this star were 
taken in 1997. The strongly magnetic star HD\,70702, with a magnetic field 
modulus of about 15\,kG, was selected in our observations as a standard star 
to check the proper technical functionality of the analysing optics of 
HARPS\-pol during our observing run in 2019. Observations of this star at 
two different rotation phases indicate strong changes in the magnetic field 
strengths and variable spectral appearance. Strongly magnetic stars are of 
special interest because only in these stars the effect of the magnetic 
field on the stellar atmosphere can be studied in great detail. Since 
magnetism affects atomic diffusion 
\citep[e.g.][]{AlecianStift},
the obtained information on the magnetic field  geometry is frequently used 
to study the horizontal accumulation and vertical stratification of 
abundances of various chemical elements 
\citep[e.g.][]{Hubrig2018,silva2020}.

In the following sections we present the acquired HARPS\-pol observations, 
their reduction and the analysis, describe the measurement results for each 
individual target, and discuss their usefulness for a better understanding 
of stellar magnetism in Ap and Bp stars. All available TESS data are 
also used to identify any significant periodicities.


\section{Observations}
\label{sec:obs}

\subsection{HARPS\-pol observations}

\begin{figure}
\centering 
\includegraphics[width=0.48\textwidth]{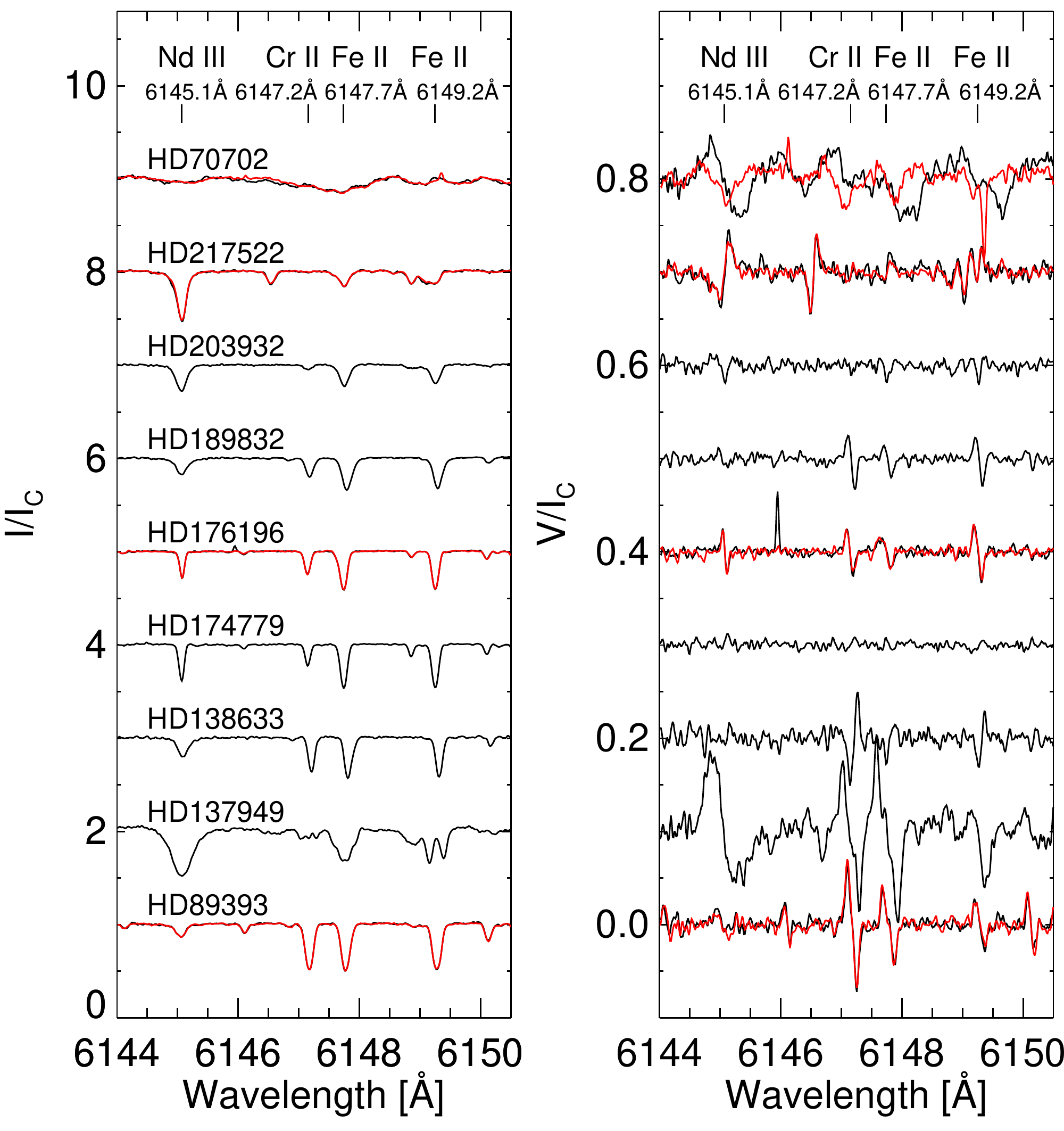}
\caption{
HARPS\-pol Stokes~$I$ (left side) and Stokes~$V$ (right side) spectra in the 
spectral region containing spectral lines characteristic for Ap stars. 
The spectra are overplotted for stars which were observed twice. The 
identified lines are \ion{Nd}{iii} at 6145.1\,\AA, \ion{Cr}{ii} at 
6147.2\,\AA, \ion{Fe}{ii} at 6147.7\,\AA, and \ion{Fe}{ii} at 6149.2\,\AA.
All spectra are shifted to the laboratory wavelengths for comparison reasons.
}
\label{fig:region}
\end{figure}

\begin{table*}
\caption{
The logbook of observations and the results of the magnetic field 
measurements for the investigated stars based on line lists that include 
all lines. The first column gives the name of the star followed by the 
spectral classification from
\citet{Renson}.
The third column presents the MJD values at the middle of the exposure, 
while the fourth column presents the signal to noise ratio measured in the 
spectral region shown in Fig.~\ref{fig:region}. The remaining columns 
show the total number of lines in each line mask, the average effective 
Land\'e factor $\overline{g}_{\rm eff}$ calculated for each line mask, the 
measured LSD mean longitudinal magnetic field strength, the false alarm 
probability (FAP) value for each measurement, and the detection flag, where 
DD means definite detection, MD marginal detection, and ND no detection.
\label{tab:obsall}
}
\centering
\begin{tabular}{lllr lr r@{$\pm$}l cc}
\hline
\multicolumn{1}{c}{Object} &
\multicolumn{1}{c}{Spectral} &
\multicolumn{1}{c}{MJD} &
\multicolumn{1}{c}{$S/N$} &
\multicolumn{1}{c}{N. of} &
\multicolumn{1}{c}{$\overline{g}_{\rm eff}$} &
\multicolumn{2}{c}{$\left< B_{\rm z}\right>$} &
\multicolumn{1}{c}{FAP} &
\multicolumn{1}{c}{Det.} \\
\multicolumn{1}{c}{} &
\multicolumn{1}{c}{Type} &
\multicolumn{1}{c}{} &
\multicolumn{1}{c}{} &
\multicolumn{1}{c}{lines} &
\multicolumn{1}{c}{} &
\multicolumn{2}{c}{(G)} &
\multicolumn{1}{c}{} &
\multicolumn{1}{c}{flag} \\
\hline
HD\,89393  & A0\,SrCrEu & 58\,646.094 & 434 & 86  & 1.20 & 314    & 7   & $<10^{-10}$      & DD \\
           &            & 58\,649.023 & 351 &     &      & 321    & 6   & $<10^{-10}$      & DD \\
HD\,137949 & F0\,SrEuCr & 56\,145.020 & 175 & 173 & 1.18 & 1543   & 19  & $<10^{-10}$      & DD \\
HD\,138633 & F0\,SrEuCr & 58\,650.051 & 239 & 116 & 1.21 & $-$205 & 9   & $<10^{-10}$      & DD \\
HD\,174779 & A0\,Si     & 58\,647.316 & 324 & 80  & 1.21 & $-$45  & 3   & $<10^{-10}$      & DD \\
HD\,176196 & B9\,EuCr   & 58\,648.375 & 309 & 96  & 1.21 & 120    & 18  & $<10^{-10}$      & DD \\
           &            & 58\,650.344 & 402 &     &      & 109    & 17  & $<10^{-10}$      & DD \\
HD\,189832 & A6\,SrCrEu & 58\,647.406 & 311 & 141 & 1.22 & 143    & 4   & $<10^{-10}$      & DD \\
HD\,203932 & A5\,SrEu   & 58\,648.430 & 290 & 141 & 1.20 & 21     & 4   & $<10^{-10}$      & DD \\
HD\,217522 & A5\,SrEuCr & 56\,148.297 & 218 & 120 & 1.24 & $-$401 & 6   & $<10^{-10}$      & DD \\
           &            & 58\,647.438 & 363 &     &      & $-$323 & 6   & $<10^{-10}$      & DD \\
HD\,70702  & B9\,EuCrSr & 58\,648.008 & 188 & 72  & 1.19 & 4388   & 169 & $<10^{-10}$      & DD \\
           &            & 58\,650.973 & 319 &     &      & 1233   & 102 & $<10^{-10}$      & DD \\
\hline
\end{tabular}
\end{table*}


All spectropolarimetric observations used in our work were acquired with 
HARPS\-pol on the ESO 3.6-m telescope on La Silla, which has a resolving 
power of $R=115\,000$ and a wavelength coverage from 3780 to 6910\,\AA{}, with 
a small gap between 5259 and 5337\,\AA. Observations were carried out in 
2019 June (Prg.~ID 0103.C-0240). Additionally, we discuss archival 
observations for HD\,137949 and HD\,217522, which were obtained in 2012 August 
(Prg.~ID 089.D-0383). The data reduction was carried out on La Silla using 
the HARPS data reduction pipeline. The normalization of the spectra to the 
continuum level is described in detail by 
\citet{Hubrig2013}.
More details on the observations are presented in Table~\ref{tab:obsall}. For 
each star in our sample, HARPS\-pol Stokes~$I$ and Stokes~$V$ spectra in the 
spectral region containing spectral lines characteristic for Ap stars are 
presented in Fig.~\ref{fig:region}.

\subsection{UVES and CES}

To check the short- and long-term line profile variability of HD\,138633, we 
extracted from the ESO archive high-resolution observations obtained with 
UVES (the Ultraviolet and Visual Echelle Spectrograph) mounted on Unit 
Telescope\,2 (UT2) of the Very Large Telescope (VLT) at Cerro Paranal, Chile 
(Prg.~ID 072.D-013, carried out on 2004 March 5), as well as two 
observations with the CES Very Long Camera, which was previously installed 
at the 1.4\,m CAT telescope (Prg.~ID 68.D-0445, carried out on 2002 
January 25). The original UVES data consist of a large amount of very short, 
noisy exposures covering the wavelength range 4959--7071\,\AA{} and a 
spectral resolution $R=107\,000$. The individual UVES files were combined 
into one final spectrum with $S/N=456$ using the ESO Phase 3 UVES 
pipeline\footnote{http://www.eso.org/rm/api/v1/public/releaseDescriptions/163}.
Wavelength calibrations for the CES spectra were executed by utilising the 
Th-Ar comparison spectra obtained immediately before and after recording the 
stellar spectrum. The first spectrum has an exposure time of 600\,s,
whereas the second spectrum was obtained 14 minutes later with an exposure
time of 780\,s.


\subsection{TESS photometry}

For three stars in our sample, HD\,176196, HD\,203932, and HD\,217522,
an analysis of TESS photometry was presented in the past. No 
variability was detected by 
\citet{MathysTESS} 
in the available TESS data for HD\,176196 and HD\,217522. Both stars were 
classified by these authors as super slowly rotating Ap (ssrAp) stars.
\citet{Cunhaetal} and \citet{Hold2021} 
analysed TESS data for HD\,203932 and reported a rotation period of 
$6.44\pm0.01$\,d, and pulsation periods of 2.6985 and 2.8048 mHz, confirming 
that this star is a rapidly oscillating Ap (roAp) star.

Three stars -- HD\,70702, HD\,89393, and HD\,174779 -- were observed during 
Year~1 of the TESS mission, in sectors~8 and 9, sector~9, and sector~13, 
respectively\footnote{These observations occurred between 2018 July 
to 2019 July.}, but no analysis of the data for these stars has been 
reported in the past. In our study, the 2-minute cadence TESS observations 
in sectors 8--9 are used to search for periodicity in the light curve of 
HD\,70702. HD\,89393 has 2-min cadence data from sector 9 and 10-min full 
frame image (FFI) data from sector 36, while HD\,174779 has 2-minute cadence 
data from sector 13. No TESS data exist for HD\,137949, but it was observed 
as a part of the Kepler 
\citep{kepler}
K2 
\citep{k2}
Campaign~15, from mid- to late-2017, in 1-minute cadence. In addition, we 
discuss the analysis of TESS data of HD\,138633, which was observed in 
sector 51 in 2022 May. We do not have TESS data for HD\,189832, which will 
be observed in sector 67, probably in 2023 July.

The shorter-cadence data from both Kepler and TESS are available in both 
simple aperture photometry (SAP) and presearch data conditioning SAP 
(PDCSAP) forms. Data processing was done using the Science Processing 
Operations Center (SPOC) pipeline at the NASA Ames Research Center 
\citep{jenkinsSPOC2016}. 
With the PDCSAP light curves, we performed a Discrete Fourier Transform
\citep[DFT; see, e.g.][]{kurtz:dft}
to identify the dominant frequency components of the signal and their 
amplitudes.


\section{Magnetic field measurements and detected periodicities}

To increase the $S/N$ in our polarimetric measurements, we employed the 
least-squares deconvolution (LSD) technique. The details of this technique, 
as well as how Stokes~$I$ and Stokes~$V$ parameters are calculated, were 
presented by
\citet{Donati1997}.
The line masks were constructed using the Vienna Atomic Line Database
\citep[VALD3;][]{kupka2011}.
For each star, we carefully checked that the selected lines  are indeed 
present in the Stokes~$I$ spectra and do not show severe blending. The 
presence of a magnetic field is evaluated following 
\citet{Donati1992},
who defined that a Zeeman profile with a false alarm probability (FAP) 
$\leq 10^{-5}$ is considered as a definite detection (DD), 
$10^{-5} <$ FAP $\leq 10^{-3}$ as a marginal detection (MD), and 
FAP $> 10^{-3}$ as a non-detection (ND).

Previous studies of magnetic Ap and Bp stars showed that the lines of 
different elements with different abundance distributions across the stellar 
surface sample the magnetic field in different ways. Some elements, such 
as rare-earth elements (REEs), are frequently observed to concentrate close 
to the magnetic poles, whereas other elements cluster in regions closer to 
the magnetic equator. Combining lines of all elements together in the LSD 
line masks may lead to the dilution of the magnetic signal or even to its 
(partial) cancellation, if enhancements of different elements occur in 
regions of opposite magnetic polarities. Therefore, it is advisable to use 
in the measurements line masks constructed for individual elements. Among the
REEs, Pr, Nd, and Eu are well known to be concentrated in surface spots, 
whereas Fe shows a rather uniform surface distribution. Using line masks 
constructed for \ion{Fe}{i}, \ion{Fe}{ii}, \ion{Pr}{iii}, \ion{Nd}{iii}, and
\ion{Eu}{ii} for the magnetic field measurements, it is possible to get 
information about the differences in the individual surface distributions of 
these ions. We note that we do not actually know the magnetic field 
geometries of the stars in our sample, but we should still be able to 
identify the locations of surface element spots if different field strengths 
are derived by the application of the LSD technique to different line masks.
	
\begin{figure*}
\centering 
\includegraphics[width=\textwidth]{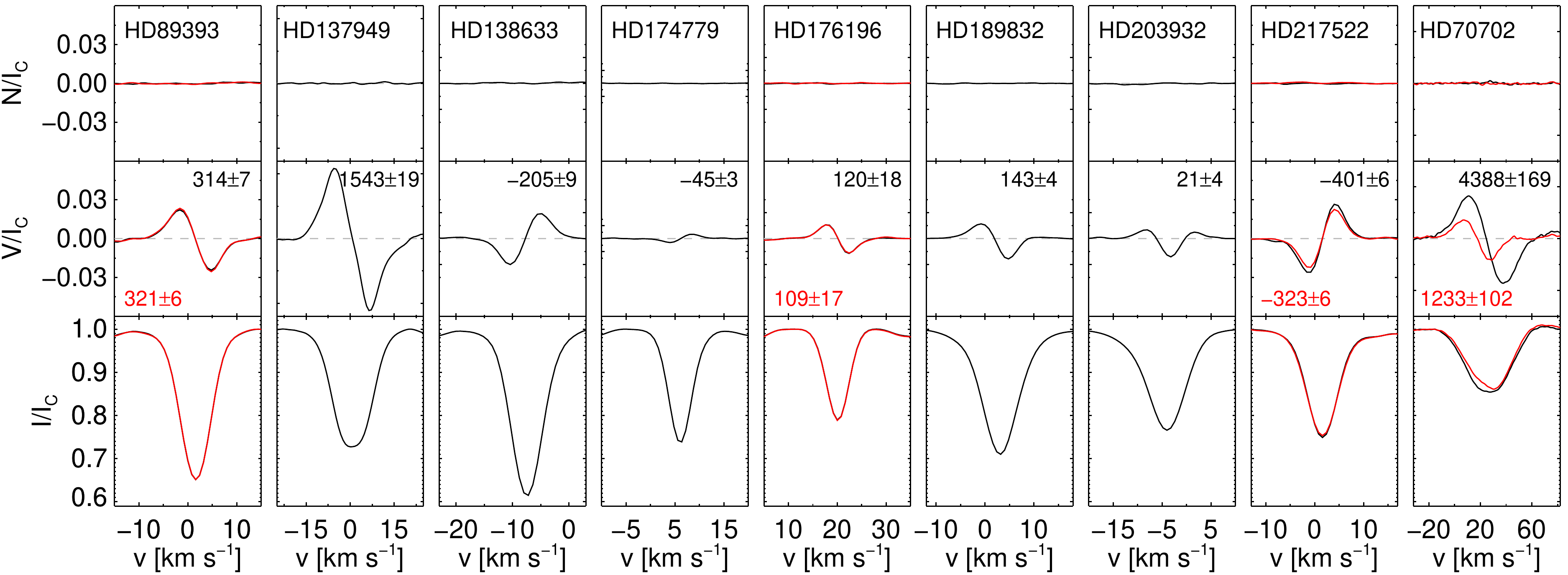}
\caption{
LSD Stokes~$I$, Stokes~$V$, and diagnostic null $N$ profiles (from bottom to 
top) calculated for the investigated stars combining line masks that include 
spectral lines belonging to different ions (\ion{Fe}{i}, \ion{Fe}{ii}, 
\ion{Pr}{iii}, \ion{Nd}{iii}, and \ion{Eu}{ii}). For stars with two 
observations, the profiles have been overplotted, and the profiles from the 
latter epoch are shown in red. In all plots we also present the measured 
mean longitudinal magnetic field strengths, colour-coded in the same way as 
the spectra.
}
\label{fig:IVNall}
\end{figure*}

\begin{table*}
\caption{
The LSD mean longitudinal magnetic field measurements for all targets using 
five different line masks are presented along with the MJD values, number of 
lines in each mask, the average Land\'e factors, and the FAP values. DD 
means definite detection, MD marginal detection, and ND no detection.
\label{tab:Bzelem}
}
\centering
\begin{tabular}{llr r@{$\pm$}l lc llr r@{$\pm$}l lc}
\hline
\multicolumn{1}{c}{MJD} &
\multicolumn{1}{c}{Mask} &
\multicolumn{1}{c}{$\overline{g}_{\rm eff}$} &
\multicolumn{2}{c}{$\left< B_{\rm z}\right>$} &
\multicolumn{1}{c}{FAP} &
\multicolumn{1}{c}{Det.} &
\multicolumn{1}{c}{MJD} &
\multicolumn{1}{c}{Mask} &
\multicolumn{1}{c}{$\overline{g}_{\rm eff}$} &
\multicolumn{2}{c}{$\left< B_{\rm z}\right>$} &
\multicolumn{1}{c}{FAP} &
\multicolumn{1}{c}{Det.}\\
\multicolumn{1}{c}{} &
\multicolumn{1}{c}{} &
\multicolumn{1}{c}{} &
\multicolumn{2}{c}{(G)} &
\multicolumn{1}{c}{} &
\multicolumn{1}{c}{flag} &
\multicolumn{1}{c}{} &
\multicolumn{1}{c}{} &
\multicolumn{1}{c}{} &
\multicolumn{2}{c}{(G)} &
\multicolumn{1}{c}{} &
\multicolumn{1}{c}{flag} \\
\hline
\multicolumn{7}{c}{HD\,89393}                                           &   \multicolumn{7}{c}{HD\,189832} \\
58\,646.094 & \ion{Fe}{i}   (49) & 1.16 & 343    & 9  & $<10^{-10}$     & DD &  58\,647.406 & \ion{Fe}{i}   (97) & 1.22 & 151    & 4  & $<10^{-10}$      & DD \\
            & \ion{Fe}{ii}  (14) & 1.17 & 262    & 14 & $<10^{-10}$     & DD &              & \ion{Fe}{ii}  (17) & 1.11 & 160    & 11 & $<10^{-10}$      & DD \\
            & \ion{Pr}{iii} (5)  & 1.11 & 246    & 20 & $<10^{-10}$     & DD &              & \ion{Pr}{iii} (10) & 0.96 & 184    & 21 & $<10^{-10}$      & DD \\
            & \ion{Nd}{iii} (6)  & 1.04 & 260    & 43 & 0.052           & ND &              & \ion{Nd}{iii} (2)  & 1.09 & 166    & 35 & 0.081            & ND \\ 
            & \ion{Eu}{ii}  (12) & 1.54 & 436    & 20 & $<10^{-10}$     & DD &              & \ion{Eu}{ii}  (15) & 1.50 & 84     & 20 & $1\times10^{-6}$ & DD \\
58\,649.023 & \ion{Fe}{i}   (49) & 1.16 & 349    & 10 & $<10^{-10}$     & DD &  \multicolumn{7}{c}{HD\,203932} \\
            & \ion{Fe}{ii}  (14) & 1.17 & 240    & 16 & $<10^{-10}$     & DD &  58\,648.430 & \ion{Fe}{i}   (98) & 1.21 & 3      & 5  & $<10^{-10}$      & DD \\\
            & \ion{Pr}{iii} (5)  & 1.11 & 292    & 18 & $<10^{-10}$     & DD &              & \ion{Fe}{ii}  (10) & 1.17 & 38     & 13 & $<10^{-10}$      & DD \\
            & \ion{Nd}{iii} (6)  & 1.04 & 249    & 39 & 0.026           & ND &              & \ion{Pr}{iii}  (8) & 0.92 & 99     & 42 & $<10^{-10}$      & DD \\
            & \ion{Eu}{ii}  (12) & 1.54 & 461    & 17 & $<10^{-10}$     & DD &              & \ion{Nd}{iii} (24) & 1.19 & 37     & 15 & $<10^{-10}$      & DD \\ 
\multicolumn{7}{c}{HD\,137949}                                          &                   & \ion{Eu}{ii}   (5) & 1.65 & 150    & 19 & $<10^{-10}$      & DD \\
56\,145.020 & \ion{Fe}{i}   (98) & 1.20 & 1548   & 32 & $<10^{-10}$     & DD &  \multicolumn{7}{c}{HD\,217522} \\
            & \ion{Fe}{ii}  (19) & 1.16 & 1291   & 55 & $<10^{-10}$     & DD &  56\,148.297 & \ion{Fe}{i}   (80) & 1.23 & $-$358 & 7  & $<10^{-10}$      & DD \\
            & \ion{Pr}{iii} (34) & 1.03 & 1941   & 48 & $<10^{-10}$     & DD &              & \ion{Fe}{ii}   (9) & 1.10 & $-$429 & 36 & $<10^{-10}$      & DD \\
            & \ion{Nd}{iii} (14) & 1.24 & 1091   & 38 & $<10^{-10}$     & DD &              & \ion{Pr}{iii}  (6) & 1.07 & $-552$ & 59 & $<10^{-10}$      & DD \\
            & \ion{Eu}{ii}  (8)  & 1.54 & 1368   & 30 & $<10^{-10}$     & DD &              & \ion{Nd}{iii} (20) & 1.30 & $-$343 & 6  & $<10^{-10}$      & DD \\
\multicolumn{7}{c}{HD\,138633}                                          &                   & \ion{Eu}{ii}   (5) & 1.60 & $-$423 & 8  & $<10^{-10}$      & DD \\
58\,650.051 & \ion{Fe}{i}   (84) & 1.21 & $-$204 & 10 & $<10^{-10}$     & DD &  58\,647.438 & \ion{Fe}{i}   (80) & 1.23 & $-$264 & 7  & $<10^{-10}$      & DD \\
            & \ion{Fe}{ii}  (11) & 1.07 & $-$160 & 27 & $<10^{-10}$     & DD &              & \ion{Fe}{ii}   (9) & 1.10 & $-$345 & 30 & $<10^{-10}$      & DD \\
            & \ion{Pr}{iii} (11) & 1.07 & $-$257 & 30 & $<10^{-10}$     & DD &              & \ion{Pr}{iii}  (6) & 1.07 & $-$426 & 40 & $<10^{-10}$      & DD \\
            & \ion{Nd}{iii} (4)  & 1.06 & $-$174 & 26 & $7\times10^{-4}$& MD &              & \ion{Nd}{iii} (20) & 1.30 & $-$309 & 6  & $<10^{-10}$      & DD \\
            & \ion{Eu}{ii}  (6)  & 1.75 & $-$182 & 28 & $<10^{-10}$     & DD &              & \ion{Eu}{ii}   (5) & 1.60 & $-$413 & 8  & $<10^{-10}$      & DD \\
\multicolumn{7}{c}{HD\,174779}                                          &    \multicolumn{7}{c}{HD\,70702} \\
58\,647.316 & \ion{Fe}{i}   (21) & 1.25 & $-$75  & 12 & $<10^{-9}$      & DD & 58\,648.008 & \ion{Fe}{i}   (20) & 1.26 & 4629 & 222 & $<10^{-10}$       & DD \\
            & \ion{Fe}{ii}  (14) & 1.20 & $-$11  & 3  & $<10^{-10}$     & DD &             & \ion{Fe}{ii}  (16) & 1.23 & 4144 & 153 & $<10^{-10}$       & DD \\
            & \ion{Pr}{iii} (7)  & 1.08 & $-$171 & 25 & $<10^{-10}$     & DD &             & \ion{Pr}{iii} (22) & 1.02 & 4031 & 290 & $<10^{-10}$       & DD \\
            & \ion{Nd}{iii} (35) & 1.17 & $-$69  & 5  & $<10^{-10}$     & DD &             & \ion{Nd}{iii} (8)  & 1.15 & 5788 & 467 & $<10^{-9}$        & DD \\ 
            & \ion{Eu}{ii}  (3)  & 1.62 & 5      & 53 & 0.687           & ND &             & \ion{Eu}{ii}  (6)  & 1.54 & 3223 & 183 & $<10^{-10}$       & DD \\
\multicolumn{7}{c}{HD\,176196}                                          &      58\,650.973 & \ion{Fe}{i}   (20) & 1.26 & 930  & 140 & $<10^{-10}$       & DD \\
58\,648.375 & \ion{Fe}{i}   (43) & 1.27 & 123    & 34 & $<10^{-10}$     & DD &             & \ion{Fe}{ii}  (16) & 1.23 & 833  & 74  & $<10^{-10}$       & DD \\
            & \ion{Fe}{ii}  (9)  & 1.13 & 18     & 20 & $<10^{-10}$     & DD &             & \ion{Pr}{iii} (22) & 1.02 & 1295 & 164 & $6\times10^{-7}$  & DD \\
            & \ion{Pr}{iii} (11) & 1.08 & 177    & 47 & $<10^{-10}$     & DD &             & \ion{Nd}{iii} (8)  & 1.15 & 1000 & 185 & $2\times10^{-5}$  & MD \\
            & \ion{Nd}{iii} (31) & 1.17 & 141    & 37 & $<10^{-10}$     & DD &             & \ion{Eu}{ii}  (6)  & 1.54 & 2279 & 160 & $<10^{-9}$        & DD \\
            & \ion{Eu}{ii}  (2)  & 1.61 & 129    & 68 & $8\times10^{-6}$& DD &             &                    &      & \multicolumn{2}{c}{}           & & \\
58\,650.344 & \ion{Fe}{i}   (43) & 1.27 & 96     & 33 & $<10^{-10}$     & DD &            &                     &      & \multicolumn{2}{c}{}           & & \\
            & \ion{Fe}{ii}  (9)  & 1.13 & 60     & 20 & $<10^{-10}$     & DD &            &                     &      & \multicolumn{2}{c}{}           & & \\
            & \ion{Pr}{iii} (11) & 1.08 & 188    & 61 & $<10^{-10}$     & DD &            &                     &      & \multicolumn{2}{c}{}           & & \\
            & \ion{Nd}{iii} (31) & 1.17 & 117    & 35 & $<10^{-10}$     & DD &            &                     &      & \multicolumn{2}{c}{}           & & \\
            & \ion{Eu}{ii}  (2)  & 1.61 & 90     & 77 & $5\times10^{-4}$& MD &            &                     &      & \multicolumn{2}{c}{}           & & \\
\hline
\end{tabular}
\end{table*}

The LSD longitudinal magnetic field $\left< B_{\rm z}\right>$ measurements
based on line masks that include the combined spectral lines belonging to 
different elements are listed in Table~\ref{tab:obsall}, and the 
corresponding LSD profiles are shown in Fig.~\ref{fig:IVNall}. The results 
of the LSD magnetic field measurements for all stars using individual line 
masks with the spectral lines of \ion{Fe}{i}, \ion{Fe}{ii}, \ion{Pr}{iii}, 
\ion{Nd}{iii}, and \ion{Eu}{ii} are presented in Table~\ref{tab:Bzelem}.

For each star, the results of our magnetic field measurements and the available 
information on the periodicities are discussed in the following subsections.
When studying the stars' periodicities using TESS observations, we identified a 
frequency as significant if it has a signal-to-noise ratio $S/N \geq 4.5$ 
and marginal if $3 \leq S/N < 4.5$. Any peaks with $S/N < 3$ were 
discarded as being consistent with noise and therefore merited no further 
investigation.


\subsection{HD\,89393}

\begin{figure}
\centering 
\includegraphics[width=0.48\textwidth]{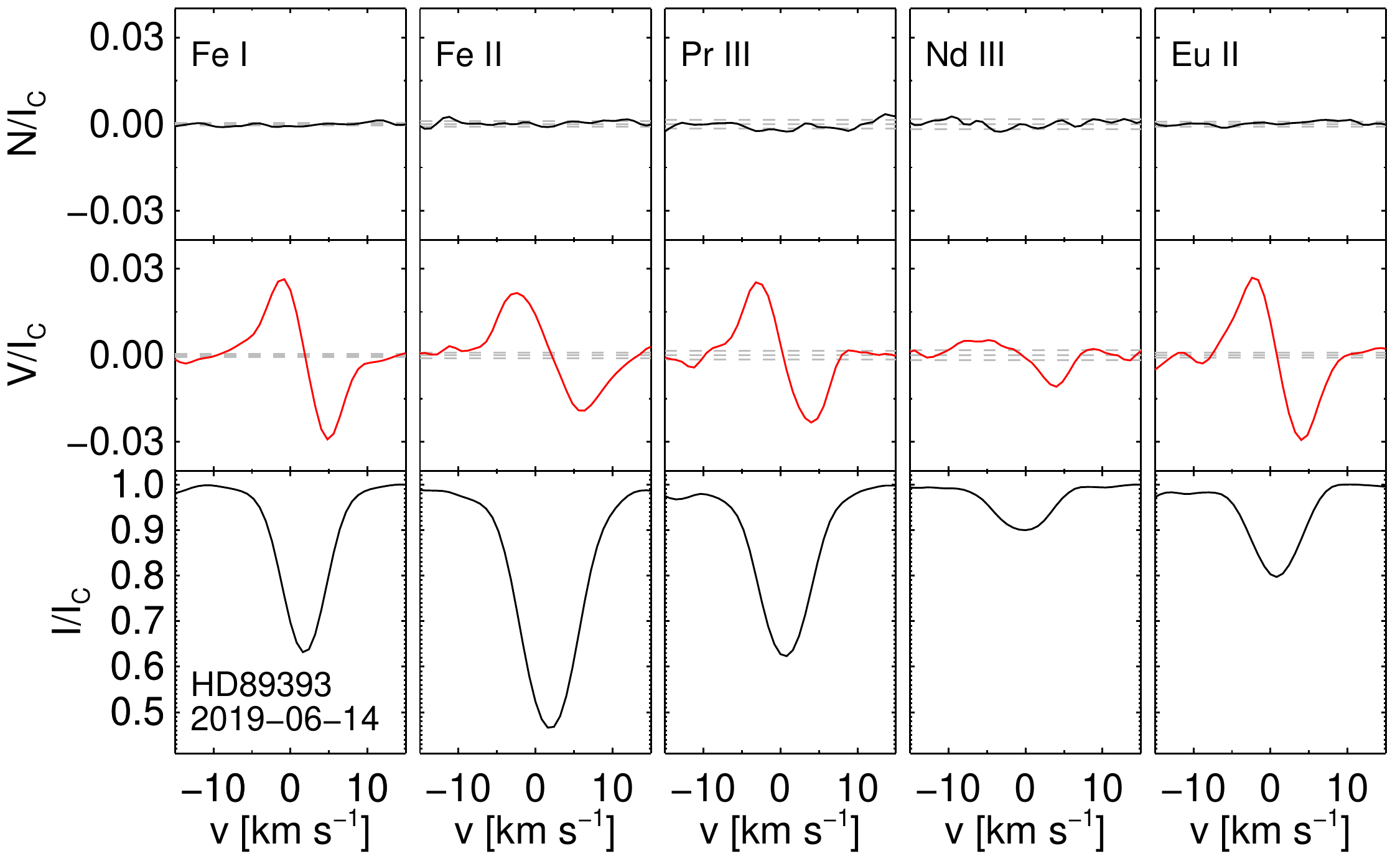}
\includegraphics[width=0.48\textwidth]{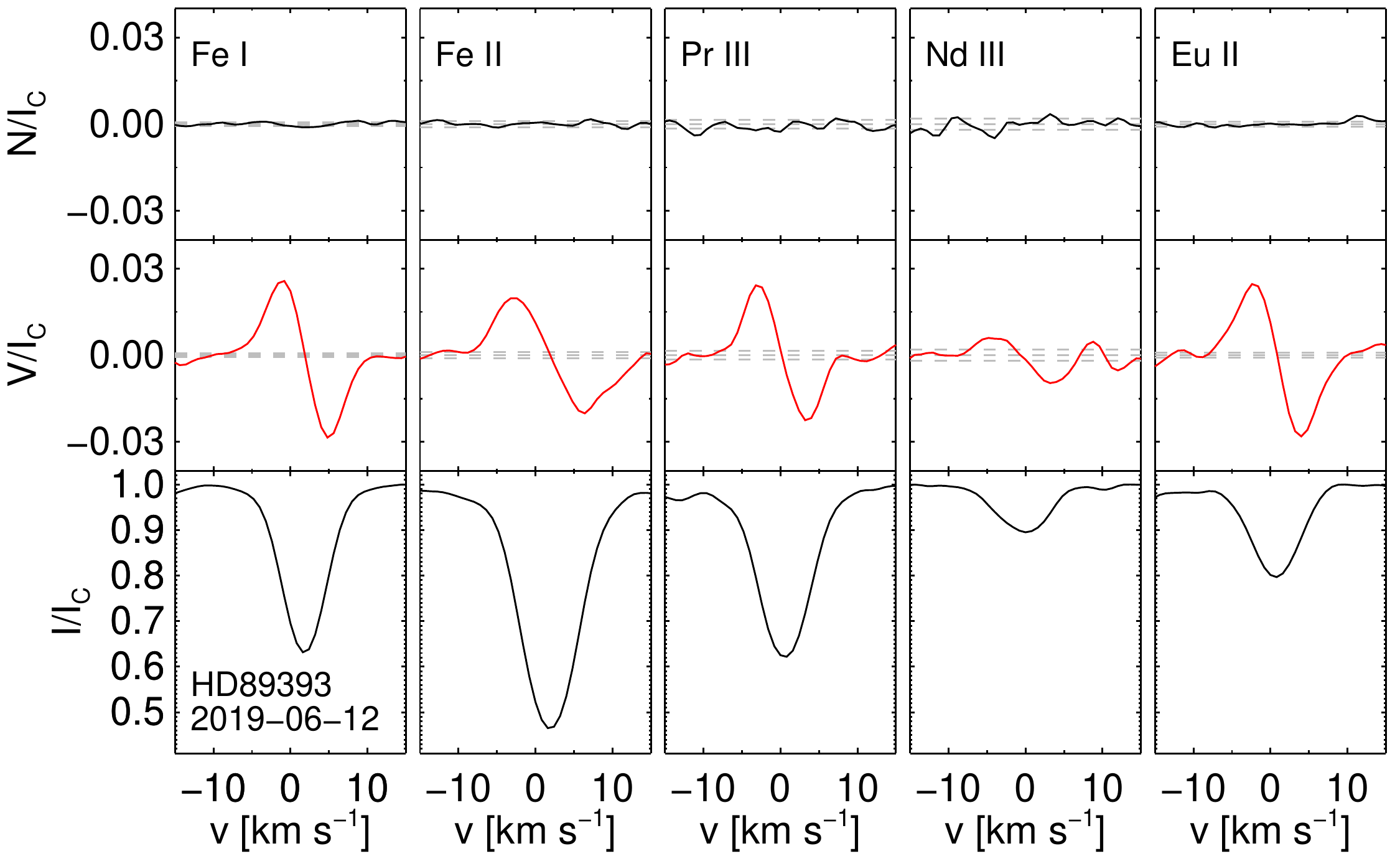}
\caption{
LSD Stokes~$I$, Stokes~$V$, and diagnostic null $N$ profiles (from bottom to 
top) calculated for the two observing epochs of HD\,89393 using line masks 
containing \ion{Fe}{i}, \ion{Fe}{ii}, \ion{Pr}{iii}, \ion{Nd}{iii}, and 
\ion{Eu}{ii} lines (from left to right).  
}
\label{fig:IVNHD89393}
\end{figure}

Not much is known about this star. It is classified as A0\,SrCrEu in
\citet{Renson}.
Magnetic field measurements were not reported for this star in the past.
In Fig.~\ref{fig:IVNHD89393}, we present the LSD Stokes~$I$, Stokes~$V$, and 
diagnostic null $N$ profiles for this star, calculated using five different 
masks. The mean longitudinal field strengths obtained at both epochs are 
almost identical (see also Tables~\ref{tab:obsall} and \ref{tab:Bzelem} and  
Fig.~\ref{fig:IVNall}), indicating the absence of any short-timescale 
variability. The amplitude of the Zeeman features and the measured field 
values clearly depend on the line masks constructed for individual elements, 
as the strongest mean longitudinal magnetic field is detected in the 
measurements using the \ion{Eu}{ii} line mask, suggesting a slightly 
different location (probably closer to the magnetic pole) of the Eu 
concentration on the stellar surface in comparison to other elements.

\begin{figure*}
\centering 
\includegraphics[width=0.9\textwidth]{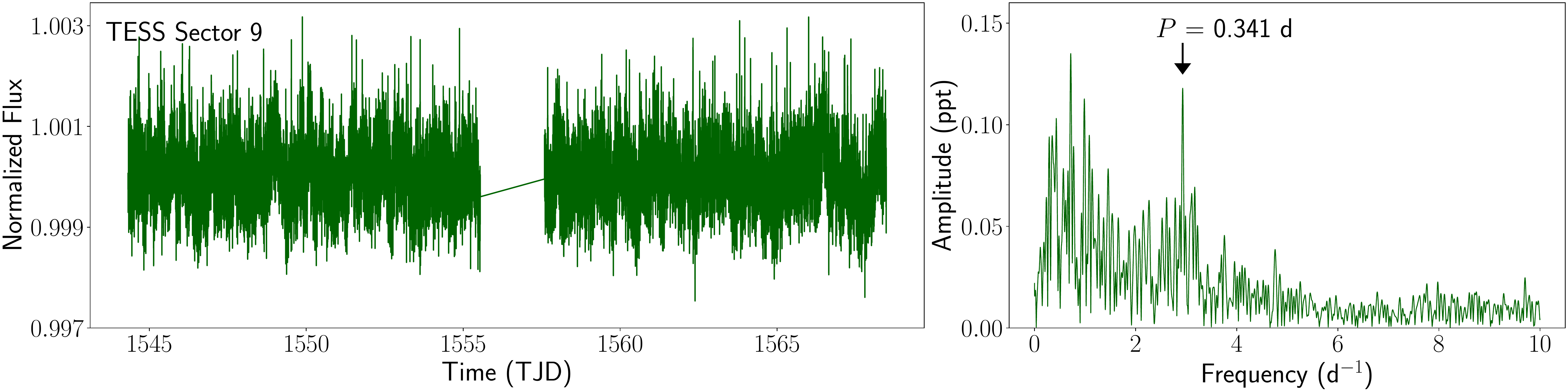}
\caption{
The left panel shows the TESS light curve of HD\,89383 from Sector 9. 
The right panel shows the periodogram of this light curve. 
The single marginal peak is highlighted with an arrow.
}
\label{fig:PHD89383}
\end{figure*}

The TESS light curve of HD\,89383 is shown in the left panel of 
Fig.~\ref{fig:PHD89383}. The \texttt{CROWDSAP} parameter suggests that there 
is significant contamination from a nearby star (TYC\,7712-2715-1). The 
\texttt{CROWDSAP} parameter refers to the fraction of flux that falls in the 
optimal aperture for the 2-minute data that is not directly attributable to 
the star in question (i.e.\ a ``crowding parameter'') and is calculated as 
part of the SPOC pipeline when data products are produced. The Gaia 
BP--RP colour index
\citep{gaia2018} 
for this nearby contaminating star is greater than 1, suggesting that it is 
a cooler star.
The periodogram shows just one marginal peak with a calculated 
$S/N$ of $\sim4$ at a period of $0.34116\pm0.00001$\,d. Since our metric is 
the relative amplitude of the peaks in the periodograms, we do not attach 
any physical interpretation to this marginal peak and are not able to 
conclude whether this frequency could represent a pulsation mode or arises 
from some other physical phenomenon, such as rotation of either HD\,89393 or 
the contaminating star. We assume that HD\,89393 likely has a long rotation 
period; this must at least be longer than the length of the TESS 
observations over 27\,d. Also the examination of the 10\,min cadence TESS 
data in sector 36 did not reveal any significant peak in our periodogram.


\subsection{HD\,137949 = 33\,Lib}

The star is classified as F0\,SrEuCr in 
\citet{Renson}.
It is a roAp star with a pulsation period of 8.3\,min  
\citep{Kurtz82}. 
The study of 
\citet{Kervella} 
indicated the presence of a companion from an analysis of proper 
motions in the Gaia Data Release~2 
\citep[GDR2;][]{gaia2018} 
and Hipparcos 
\citep{hipparcos} 
catalogues. A strong mean longitudinal magnetic field with a value of more 
than 1\,kG was initially discovered in this star by 
\citet{Babcock}.
Given the presence of the strong magnetic field and the slow rotation, 
numerous spectral lines are resolved into their magnetically split 
components. The most complete review of the previous magnetic studies of 
33\,Lib is presented in 
\citet{Mathys2017}. 
Their seven measurements of the mean magnetic field modulus 
$\left<B\right>$ using magnetically resolved lines in observations 
acquired between 1996 and 1998 appear to be constant in the range from 4.63 
to 4.69\,kG, whereas the $\left<B_{\rm z}\right>$-values from the same 
observations range from 1.47 to 1.68\,kG. Taking into account all 
previous measurements of $\left<B_{\rm z}\right>$-values, 
\citet{Mathys2017} 
suggest that this star has a rotational period $P_{\rm rot}$ of 5195\,d 
(approximately 14.25\,yr).

\begin{figure}
\centering
\includegraphics[width=0.48\textwidth]{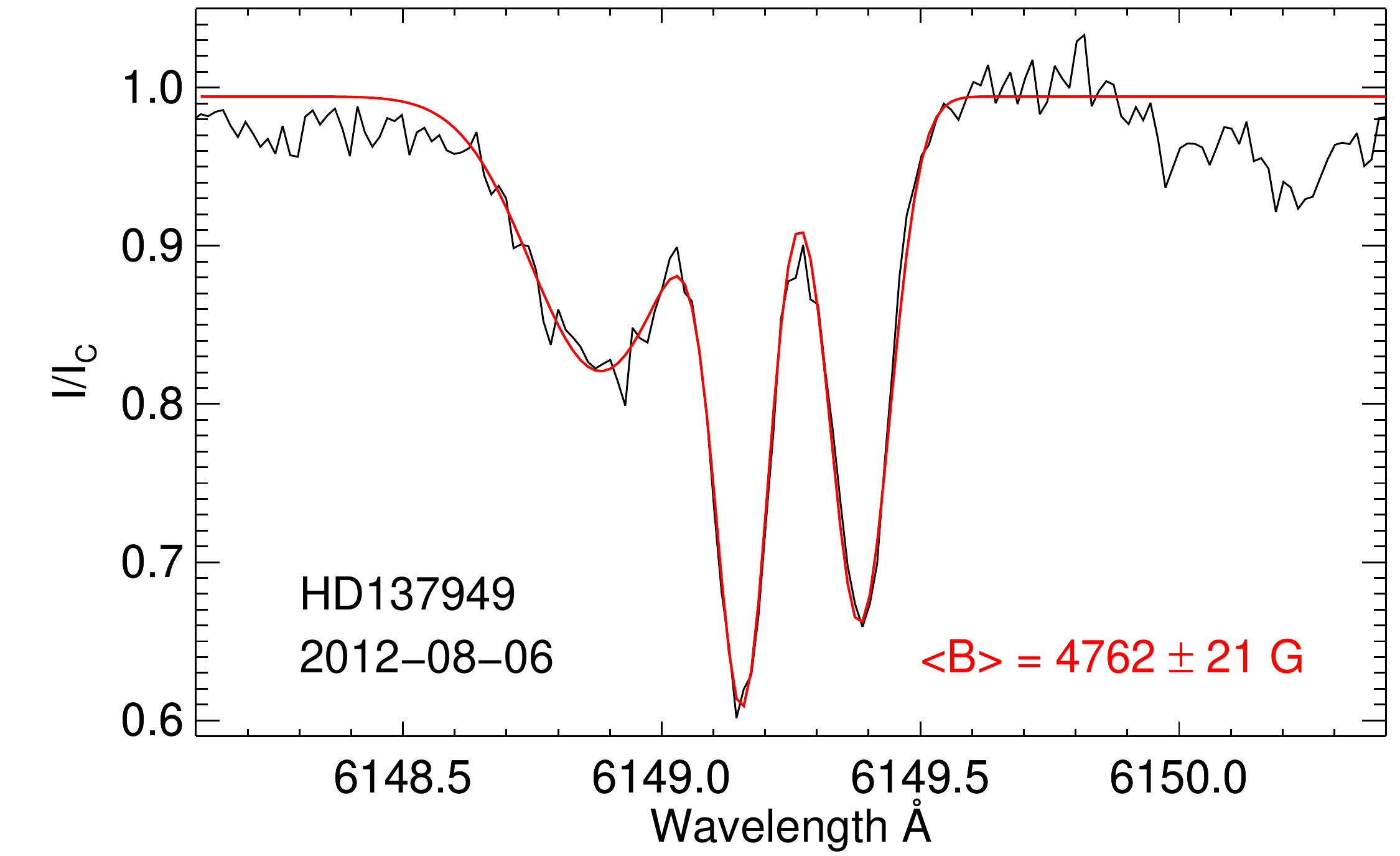}
\caption{
Stokes~$I$ spectra of HD\,137949 in the region containing the magnetically 
split \ion{Fe}{ii} line at 6149.258\,\AA. The distance between the Zeeman 
subcomponents is determined using a triple Gaussian fit represented by the 
red solid line. It has been suggested that the leftmost component corresponds
to a \ion{Sm}{iii} line.
}
\label{fig:BmodHD137949}
\end{figure}

\begin{figure}
\centering 
\includegraphics[width=0.48\textwidth]{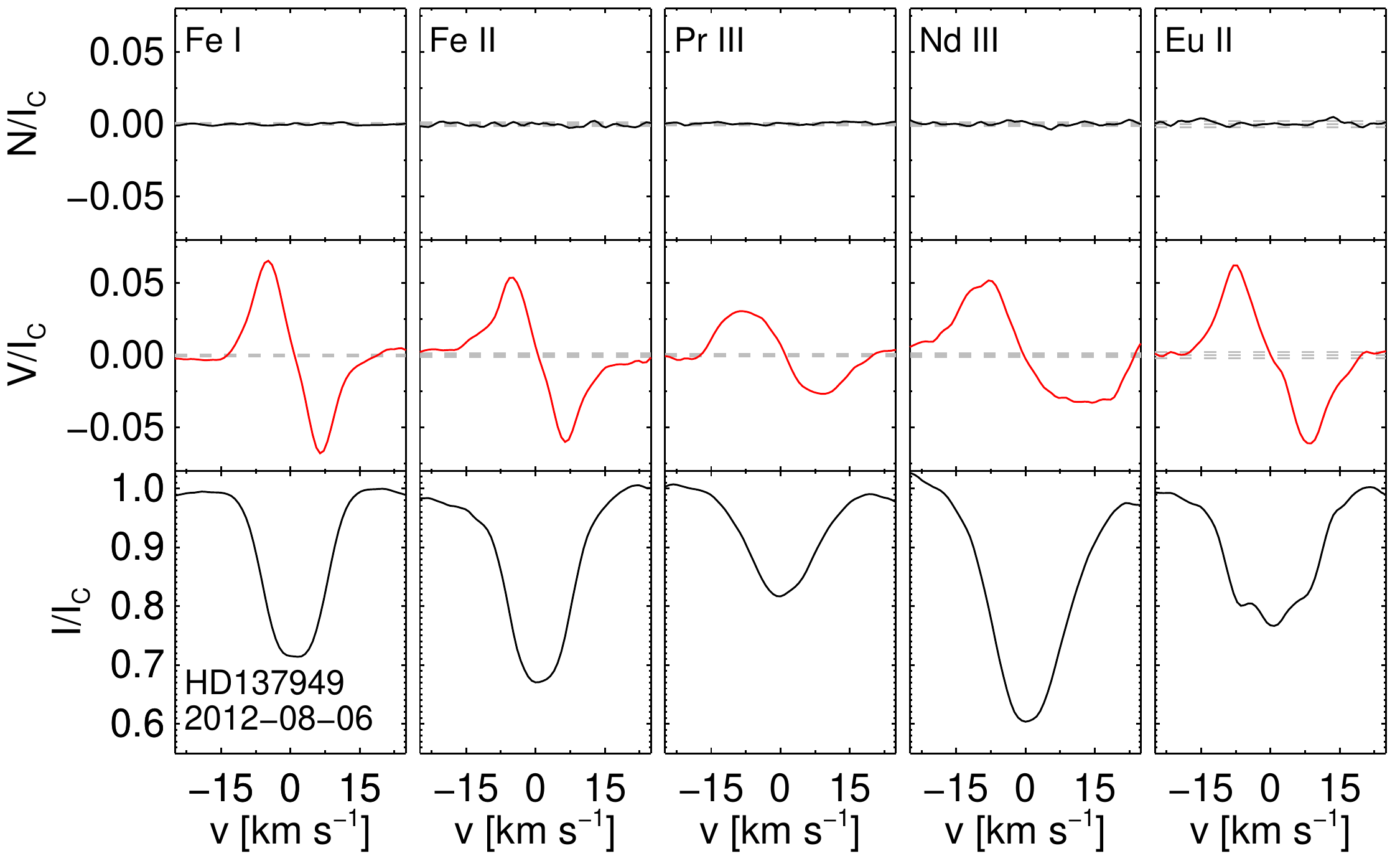}
\caption{
As Fig.~\ref{fig:IVNHD89393}, but for HD\,137949.
}
\label{fig:IVNHD137949}
\end{figure}

On the other hand, 
\citet{Giarrusso} recently reported that $\left<B\right>$ and 
$\left<B_{\rm z}\right>$-values have shown a slight increase over several 
years, indicating that the rotation period should be longer than 27\,yr. Our 
measurement of the mean field modulus using the HARPS\-pol observations
obtained in 2012 August, $\left<B\right>=4.76\pm0.02$\,kG, is of the same 
order as reported by 
\citet{Giarrusso}. 
As an illustration, we show the Zeeman subcomponents of the resolved 
magnetically split \ion{Fe}{ii} line at 6149.258\,\AA{} in 
Fig.~\ref{fig:BmodHD137949}. No measurements of the mean longitudinal 
magnetic field after 2007 were presented by 
\citet{Giarrusso}. 
Our measurement of the mean longitudinal field strength using all lines, 
$\left<B_{\rm z}\right>=1.54\pm0.02$\,kG, does not confirm the slight 
increase, but the small differences in the field values can be caused 
by the choice of spectral lines used in the measurements. As we see in 
Fig.~\ref{fig:IVNHD137949}, which presents the LSD Stokes~$I$, Stokes~$V$, 
and diagnostic null $N$ profiles calculated using five different masks, and 
in Table~\ref{tab:Bzelem}, the amplitude of the Zeeman features and the 
measured field values for this star depend on the line masks constructed 
for individual elements: We measure $\left<B_{\rm z}\right>=1.94\pm0.05$\,kG 
using 34 \ion{Pr}{iii} lines, but only 
$\left<B_{\rm z}\right>=1.09\pm0.04$\,kG using 14 \ion{Nd}{iii} lines. 
In any case, the comparison of our measurements with previous measurements 
confirm a long rotation period for HD\,137949.

A full analysis of short-period oscillations (typical for roAp stars) was done 
for HD\,137949 by
\citet{K2-HD137949}
based on 60-s cadence K2 observations. 
The authors found that the low-frequency peaks were aliases of the frequency 
splitting, and that the rotation period was likely extremely long.
The amplitude spectra of 33\,Lib were presented in 
Figures~1 and 2 of 
\citet{K2-HD137949}.


\subsection{HD\,138633}

This star is classified as F0\,SrEuCr in 
\citet{Renson}.
The presence of a weak magnetic field of the order of 0.7\,kG was
announced by 
\citet{Titarenko}. 
They report that in contrast to ordinary Ap stars that exhibit strong REE 
lines in their spectra, these elements are represented only very poorly
in the atmosphere of HD\,138633. A photometric study using the STEREO 
satellites suggested that this star is constant or probably constant
\citep{Wraight2012}. 
On the other hand, 
\citet{Romanyuk2017}
report on the change of the field polarity in the measurements of the mean 
longitudinal magnetic field in spectra acquired in 2010 on two different 
nights separated by 5 days, with $\bz=310\pm30$\,G and $\bz=-290\pm20$\,G.

\begin{figure}
\centering 
\includegraphics[width=0.48\textwidth]{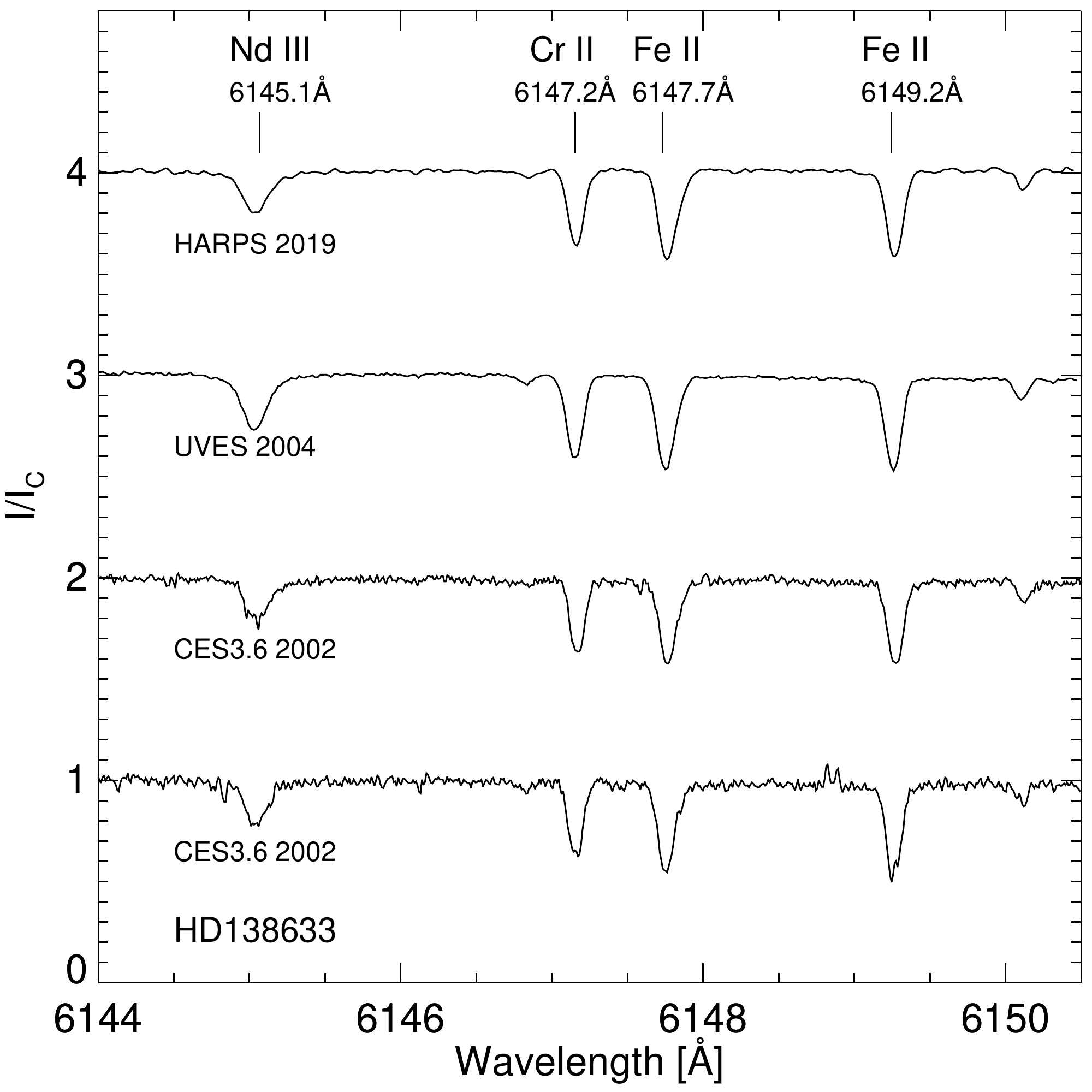}
\caption{
The same spectral region as shown in Fig.~\ref{fig:region} but showing the 
spectra for HD\,138633 obtained with different instruments between the years 
2002 and 2019. The second CES spectrum was obtained 14 minutes after 
the first spectrum on 2002 January 25.
}
\label{fig:HD138633reg}
\end{figure}

To check the short- and long-term line profile variability of HD\,138633, we 
compared the line profile shapes observed with HARPS\-pol in 2019 with those 
observed using UVES on 2004 March 5 and CES on 2002 January 25. In 
Fig.~\ref{fig:HD138633reg}, we display a few line profiles observed with 
these spectrographs in the same spectral region as shown in  
Fig.~\ref{fig:region}. Some tiny changes in the line shapes between the two 
CES spectra separated in time by 14 minutes seem to exist, but obviously 
higher $S/N$ data are needed to achieve any conclusion about any rapid 
spectral variability of this star.

\begin{figure}
\centering 
\includegraphics[width=0.48\textwidth]{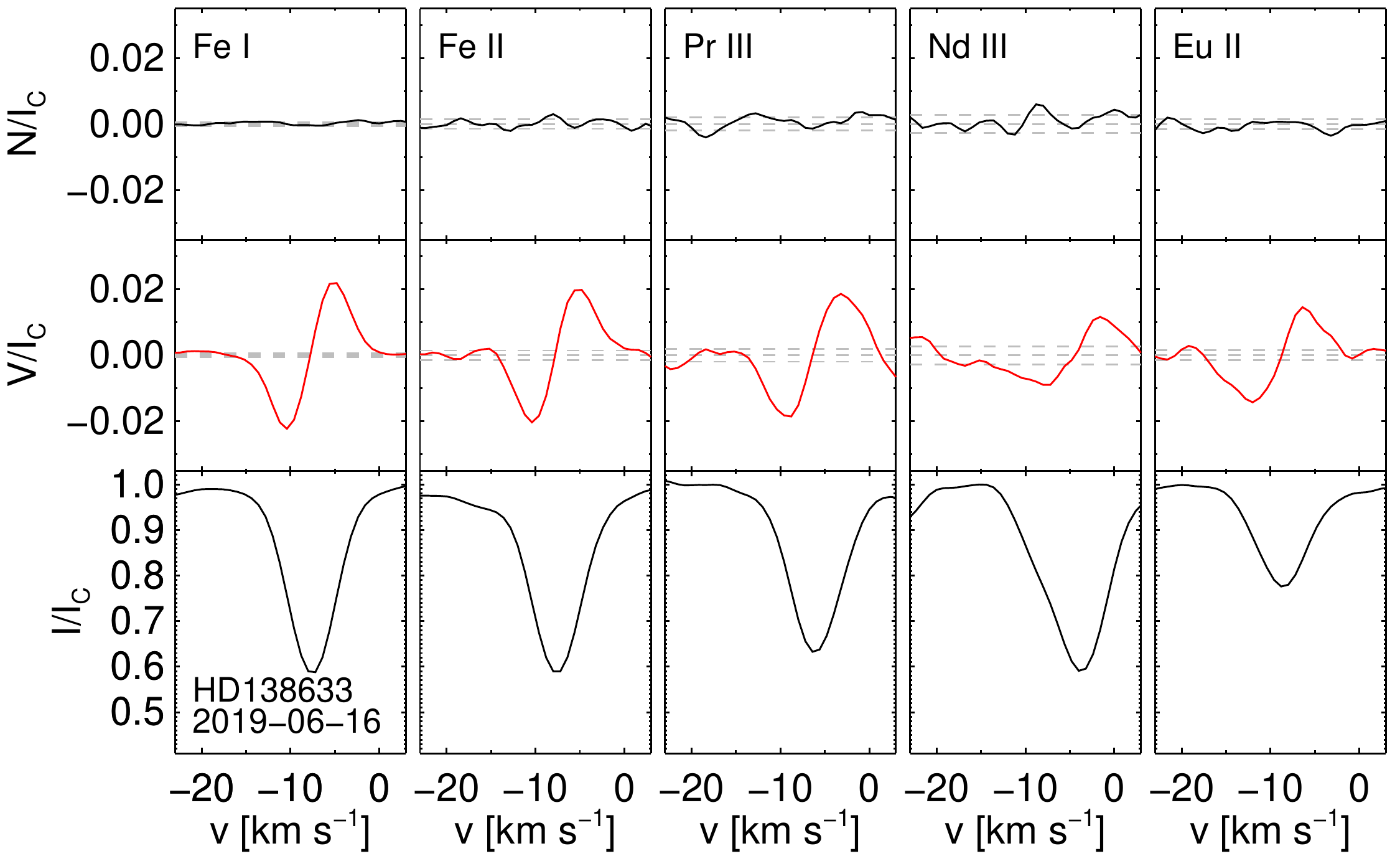}
\caption{
As Fig.~\ref{fig:IVNHD89393} but for HD\,138633.
}
\label{fig:IVNHD138633}
\end{figure}

In Fig.~\ref{fig:IVNHD138633} we present the LSD Stokes~$I$, Stokes~$V$, and 
diagnostic null $N$ profiles calculated using five different masks. Using 
all lines for the measurements, we obtain a positive longitudinal magnetic 
field on the order of 200\,G, which is comparable to measurements carried 
out using different line masks (see Tables~\ref{tab:obsall}--\ref{tab:Bzelem}). 
The strongest field, $\left<B_{\rm z}\right>=257\pm30$\,G, is detected for 
the \ion{Pr}{iii} line mask. Based on the strengths of the LSD Stokes~$I$ 
profiles of the studied REEs, we cannot confirm the finding of 
\citet{Titarenko} 
that these elements are represented very poorly in the atmosphere 
of HD\,138633.

\begin{figure*}
\centering
\includegraphics[width=0.9\textwidth]{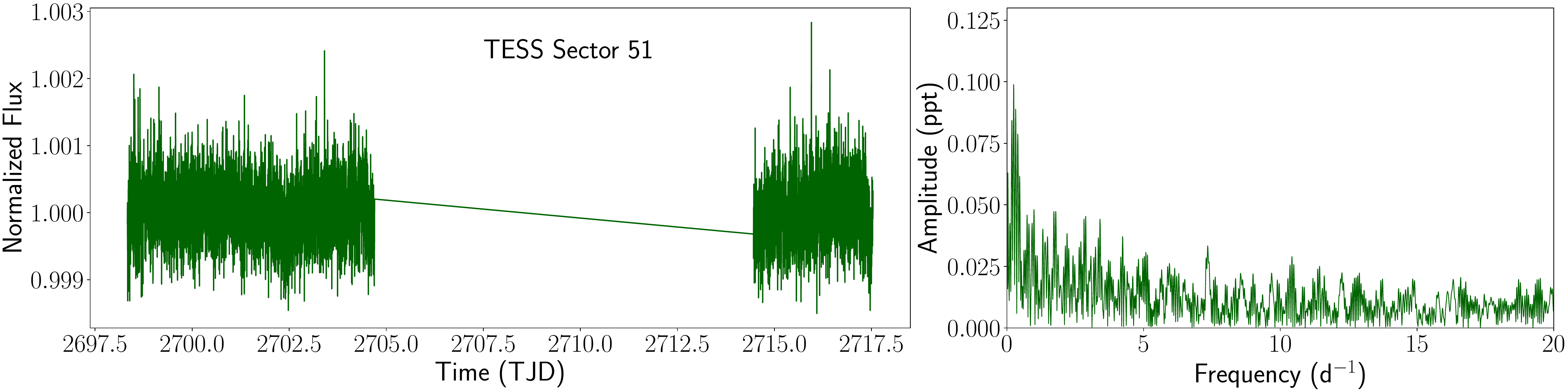}
\caption{
The left panel shows the TESS light curve of HD\,138633 from Sector 51.
The right panel shows the periodogram of this light curve. No significant 
peaks are detected; the higher-amplitude peaks at low frequencies are likely 
due to red noise and the poor quality of the data -- i.e., its short length 
(much of the data was cut due to contamination by scattered light from the 
Earth and the Moon). 
}
\label{fig:PHD138633}
\end{figure*}

Due to significant scattered light from both the Moon and the Earth entering 
Camera 1, the short-cadence 
TESS data of HD\,138633
suffers from a data gap of approximately 10 days. Using the available
photometry, we performed a DFT (see, Fig.~\ref{fig:PHD138633}) and were 
unable to find any significant peaks in the periodogram that could 
correspond to either pulsations or rotation. This suggests that this 
star has a long rotational period that is at least longer than the length of 
the TESS observations. Due to the short length and poor quality of the TESS 
short-cadence data, we are unable to identify any evidence for the finding of 
\citet{Titarenko} 
that this star belongs to the group of rapidly oscillating Ap stars and 
possesses a pulsation period of $\sim17$\,min.


\subsection{HD\,174779}

\begin{figure}
\centering 
\includegraphics[width=0.48\textwidth]{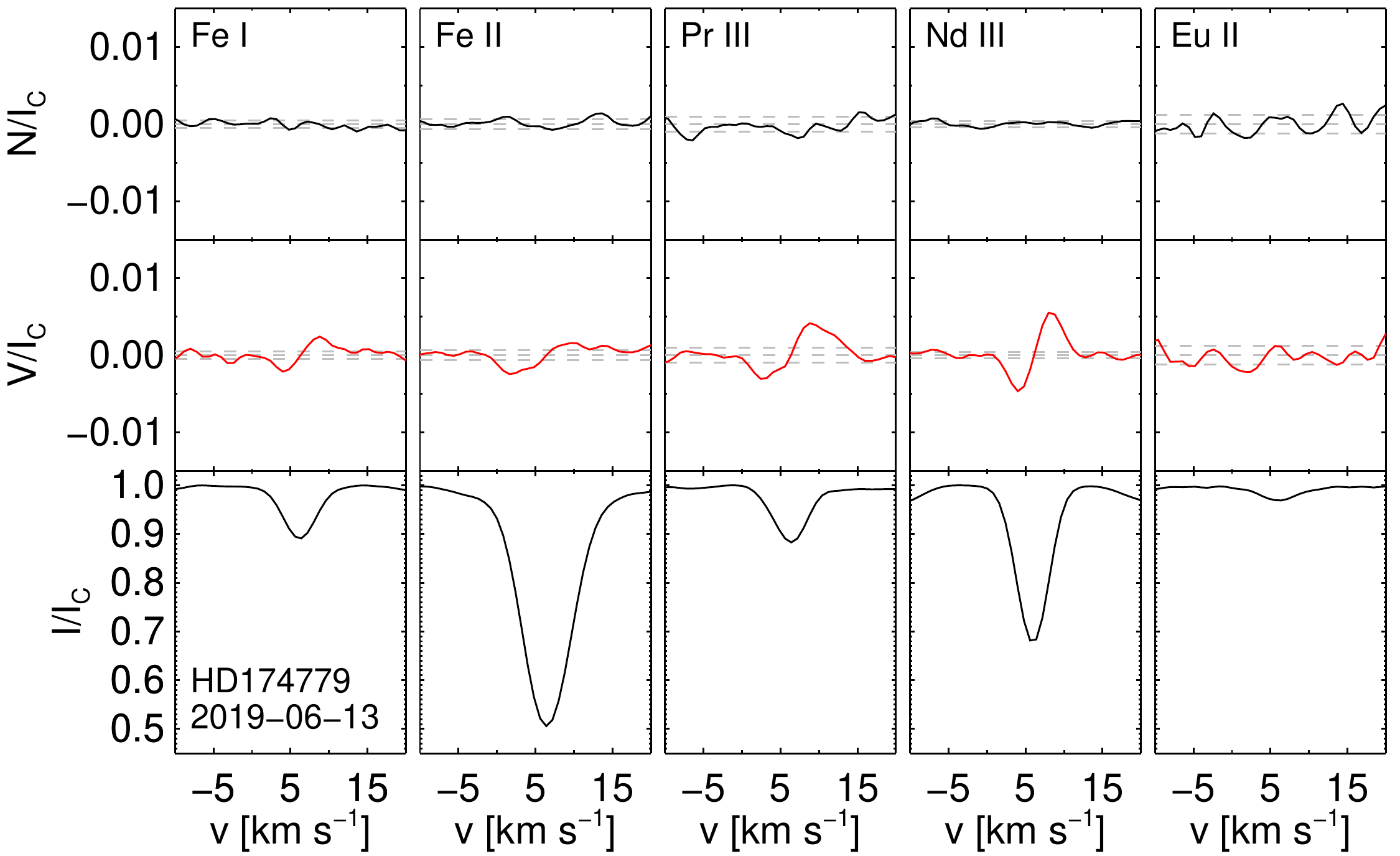}
\caption{
As Fig.~\ref{fig:IVNHD89393}, but for HD\,174779.
}
\label{fig:IVNHD174779}
\end{figure}

The star is classified as A0\,Si in 
\citet{Renson}.
\citet{CatRen} 
noted that it exhibits variability in its luminosity and/or colour, but do 
not report an associated period. The study of 
\citet{Kervella} 
indicated the presence of a companion from an analysis of proper motions in 
the GDR2 and Hipparcos catalogues. Magnetic field measurements have not been 
reported for this star in the past. Our measurements presented in 
Tables~\ref{tab:obsall} and \ref{tab:Bzelem} and in Figs.~\ref{fig:IVNall} 
and \ref{fig:IVNHD174779} show that the magnetic field is very weak, with 
$\left<B_{\rm z}\right>=-45\pm3$\,G measured using all lines. It is 
striking that for the \ion{Pr}{iii} line mask our measurements yield a much 
higher field strength, $\left<B_{\rm z}\right>=-171\pm25$\,G, while the 
measurements using other masks are all below $-75$\,G. These results show 
that great care should be taken in the selection of line lists for the 
measurements of weak magnetic fields.

\begin{figure*}
\centering 
\includegraphics[width=0.9\textwidth]{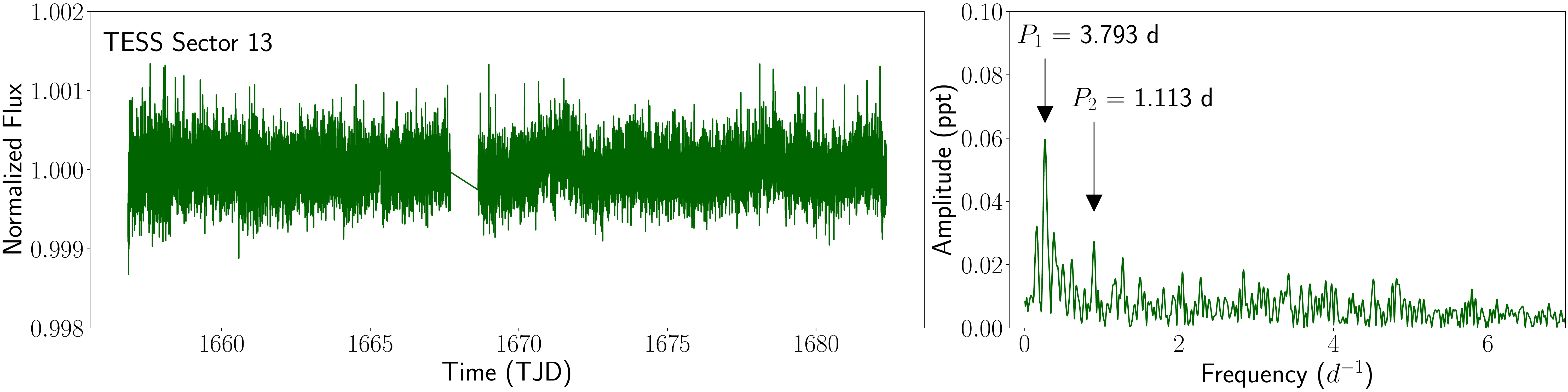}
\caption{
TESS light curve and periodogram for HD\,174779 showing two peaks -- one 
marginal (S/N of $\sim4$) low-frequency peak and another, higher-frequency 
peak with a similar S/N. 
}
\label{fig:PHD174779}
\end{figure*}

The analysis of the TESS observations of HD\,174779 reveals
two marginal peaks in the periodogram (see Fig.~\ref{fig:PHD174779}). The 
first corresponds to a period of $3.7933\pm0.0003$\,d with S/N of 4.11. The 
second, higher-frequency peak with a similar S/N corresponds to the
period of $1.11473\pm0.0001$\,d. We avoid any physical interpretation of 
these two peaks due to their marginal significance, and note that there is 
no clear, salient evidence of rotation from the light curve. 


\subsection{HD\,176196}

This star is classified as B9\,EuCr in 
\citet{Renson}. 
The study of 
\citet{Kervella} 
indicated the presence of a companion from an analysis of proper motions in 
the GDR2 and Hipparcos catalogues. First detections of the mean longitudinal 
magnetic field, $\left<B_{\rm z}\right>=258\pm69$\,G and 
$\left<B_{\rm z}\right>=174\pm58$\,G,  were reported by 
\citet{Hubrig2006} 
using low resolution spectropolarimetry with the FORS\,1 instrument 
installed at the ESO VLT.

\begin{figure}
\centering 
\includegraphics[width=0.48\textwidth]{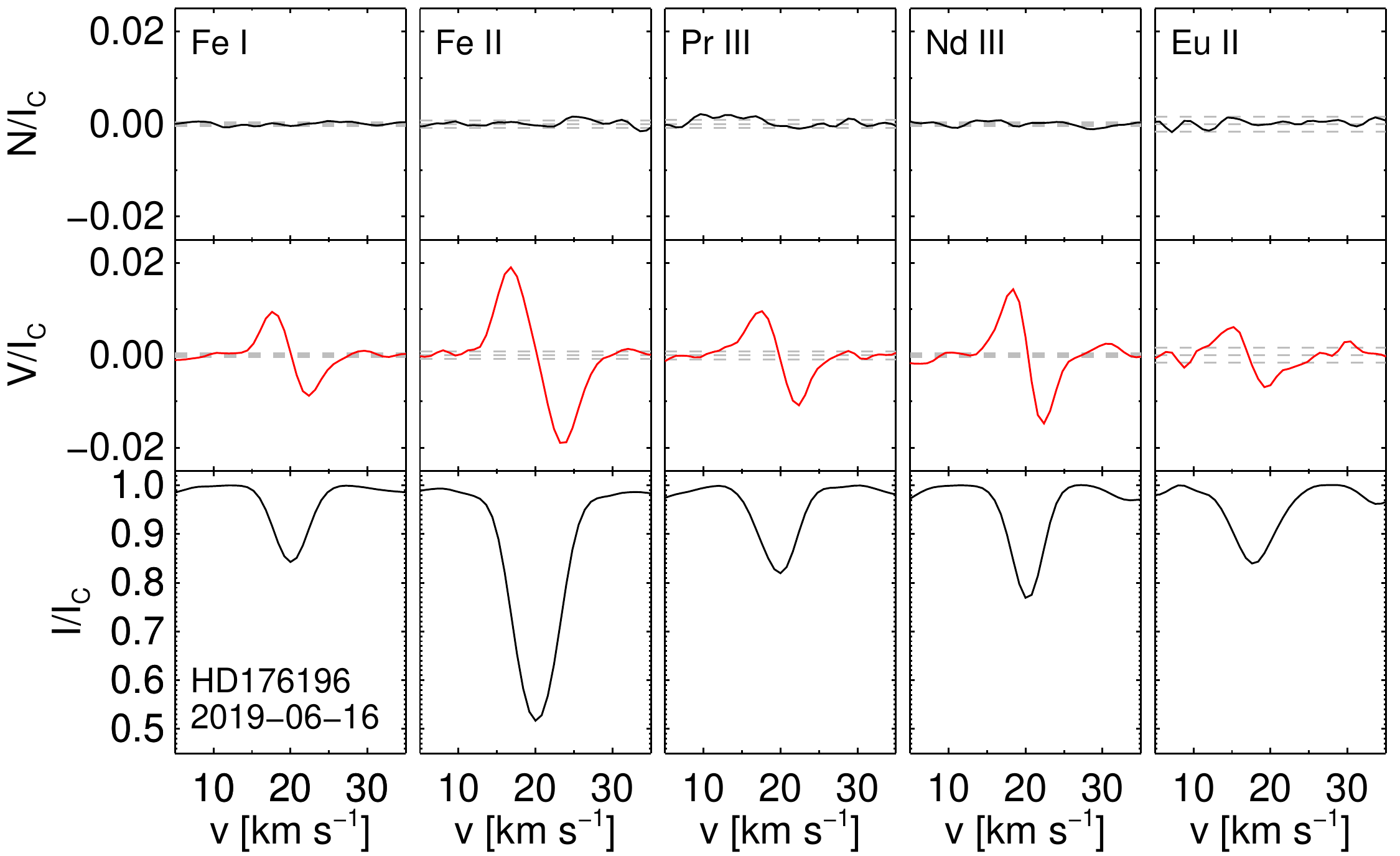}
\includegraphics[width=0.48\textwidth]{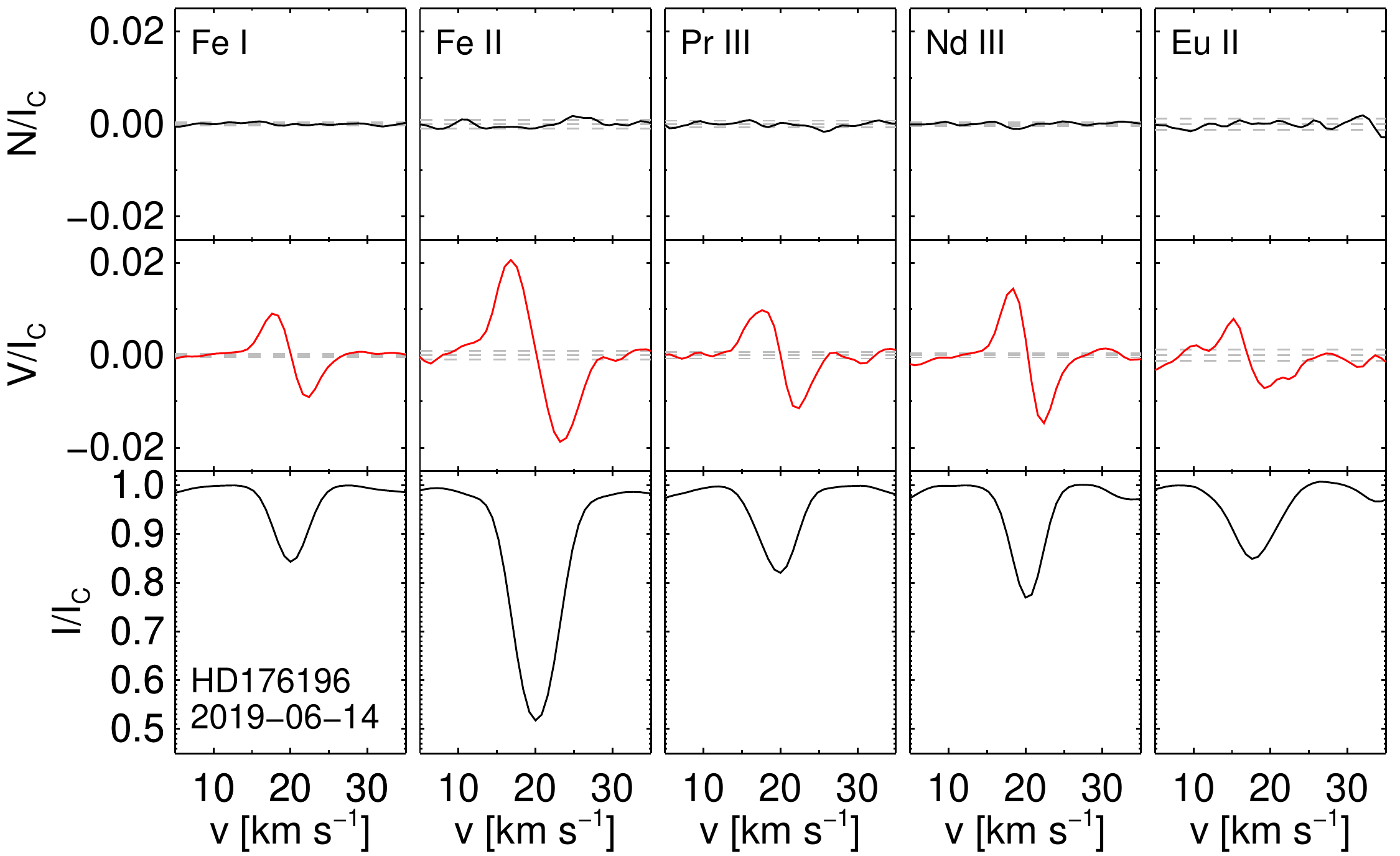}
\caption{
As Fig.~\ref{fig:IVNHD89393}, but for HD\,176196.
}
\label{fig:IVNHD176196}
\end{figure}

Our measurements, separated by two nights, are presented in 
Tables~\ref{tab:obsall} and \ref{tab:Bzelem} and in Figs.~\ref{fig:IVNall} 
and \ref{fig:IVNHD176196} and show almost identical, rather low, mean 
longitudinal field strengths on the order of 100--120\,G. For the 
measurements using the \ion{Pr}{iii} line mask, the field strength reaches 
180--190\,G. Given the small change in the field strength since the 
first observations by 
\citet{Hubrig2006},
it is very likely that the rotation period of HD\,176196 is quite long. 


\subsection{HD\,189832}

\begin{figure}
\centering 
\includegraphics[width=0.48\textwidth]{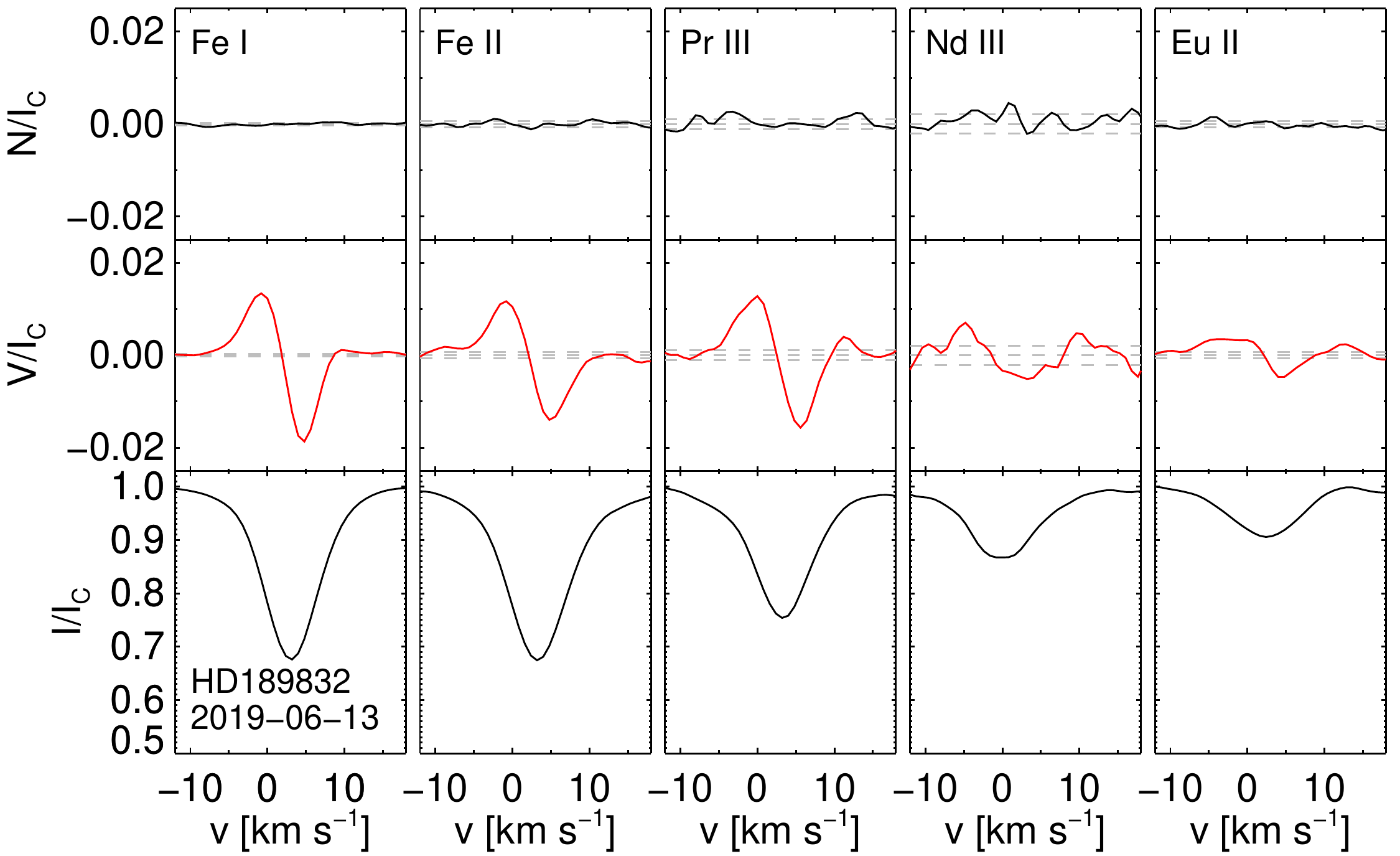}
\caption{
As Fig.~\ref{fig:IVNHD89393} but for HD\,189832.
}
\label{fig:IVNHD189832}
\end{figure}

Not much is known for this star. It is classified as A6\,SrCrEu in 
\citet{Renson}.
The study of 
\citet{Kervella} 
indicated the presence of a companion from an analysis of proper motions in 
the GDR2 and Hipparcos catalogues. A rotational period $P_{\rm rot}$ of 
18.89\,d was mentioned in 
\citet{ManMat},
who used ground-based photometry to arrive at this value. Magnetic field 
measurements have not been reported for this star in the past. Similar to 
HD\,176196, our measurements for this star, displayed in 
Tables~\ref{tab:obsall} and \ref{tab:Bzelem} and in Figs.~\ref{fig:IVNall} 
and \ref{fig:IVNHD189832}, show the presence of a positive weak magnetic 
field, with a somewhat stronger field, $\left<B_{\rm z}\right>=184\pm21$\,G, 
measured using the \ion{Pr}{iii} line mask. 


\subsection{HD\,203932}

\begin{figure}
\centering 
\includegraphics[width=0.48\textwidth]{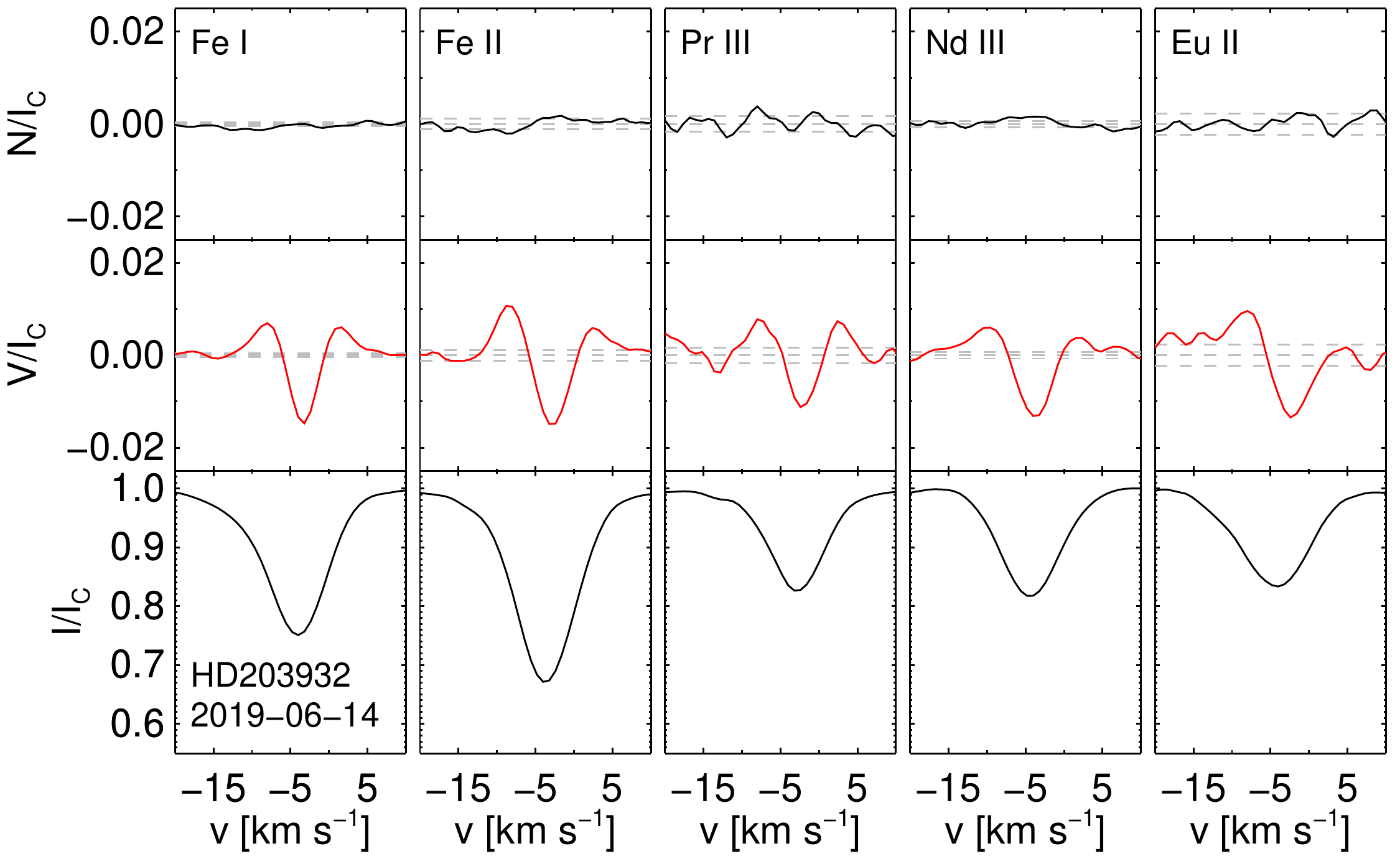}
\caption{
As Fig.~\ref{fig:IVNHD89393} but for HD\,203932.
}
\label{fig:IVNHD203932}
\end{figure}

This star is classified as A5\,SrEu in 
\citet{Renson}.
It is a roAp star 
\citep{Kurtz84}, 
with a pulsation period of about 6.2\,min. 
\citet{Hold2021} 
reported a rotational period of 6.44\,d, based on TESS data.
\citet{MatHub97} 
were unable to identify the presence of a magnetic field. Later on, 
\citet{Hubrig2004b} 
reported $\left<B_{\rm z}\right>=-267\pm72$\,G measured in low-resolution 
FORS\,1 polarimetric spectra acquired in 2002. Our analysis, shown in 
Tables~\ref{tab:obsall} and \ref{tab:Bzelem} and in Figs.~\ref{fig:IVNall} 
and \ref{fig:IVNHD203932}, yields the lowest mean longitudinal field strength 
when all lines are used in the measurement, but the field strength is as 
high as 150\,G in the measurements using only the \ion{Eu}{ii} line mask. 
Notably, the shape of the calculated Zeeman features in 
Figs.~\ref{fig:IVNall} and \ref{fig:IVNHD203932} corresponds to a typical 
crossover profile, which results from the correlation between the Zeeman 
effect and the rotation-induced Doppler effect across the stellar surface.


\subsection{HD\,217522}

This star is classified as A5\,SrEuCr in 
\citet{Renson}. 
It is a rapidly pulsating star with a pulsation period of about 13.7\,min 
\citep{Medupe}. 
\citet {Hubrig2002} 
showed that HD\,217522 is very similar to Przybylski’s star (HD\,101065), 
which exhibits the most complex spectra known, with numerous lines of 
lanthanides and also rotates extremely slowly, with a probable $P_{\rm rot}$ of 
about 188\,yr 
\citep{Hubrig2018}.
The study of 
\citet{Kervella} 
indicated the presence of a companion from an analysis of proper motions in 
the GDR2 and Hipparcos catalogues, but no companion candidate was detected 
using diffraction-limited near-infrared imaging with NAOS-CONICA at the VLT by 
\citet{Schoeller2012}.

The first detection of a magnetic field of about $-400$\,G was reported by 
\citet{MatHub97} 
using the ESO Cassegrain Echelle Spectrograph (CASPEC), fed by the ESO 3.6\,m 
telescope, for observations acquired in 1992. One additional observation was 
obtained in 1997 with $\left<B_{\rm z}\right>=-559\pm63$\,G 
\citep{Hubrig2002}.
Later, 
\citet{Hubrig2004b} 
reported $\left<B_{\rm z}\right>=-725\pm88$\,G using low-resolution 
spectropolarimetry with FORS\,1. No variability of the field strength over 
the pulsation cycle was detected by 
\citet{Hubrig2004a}
using a FORS\,1 spectropolarimetric time series.

\begin{figure}
\centering 
\includegraphics[width=0.48\textwidth]{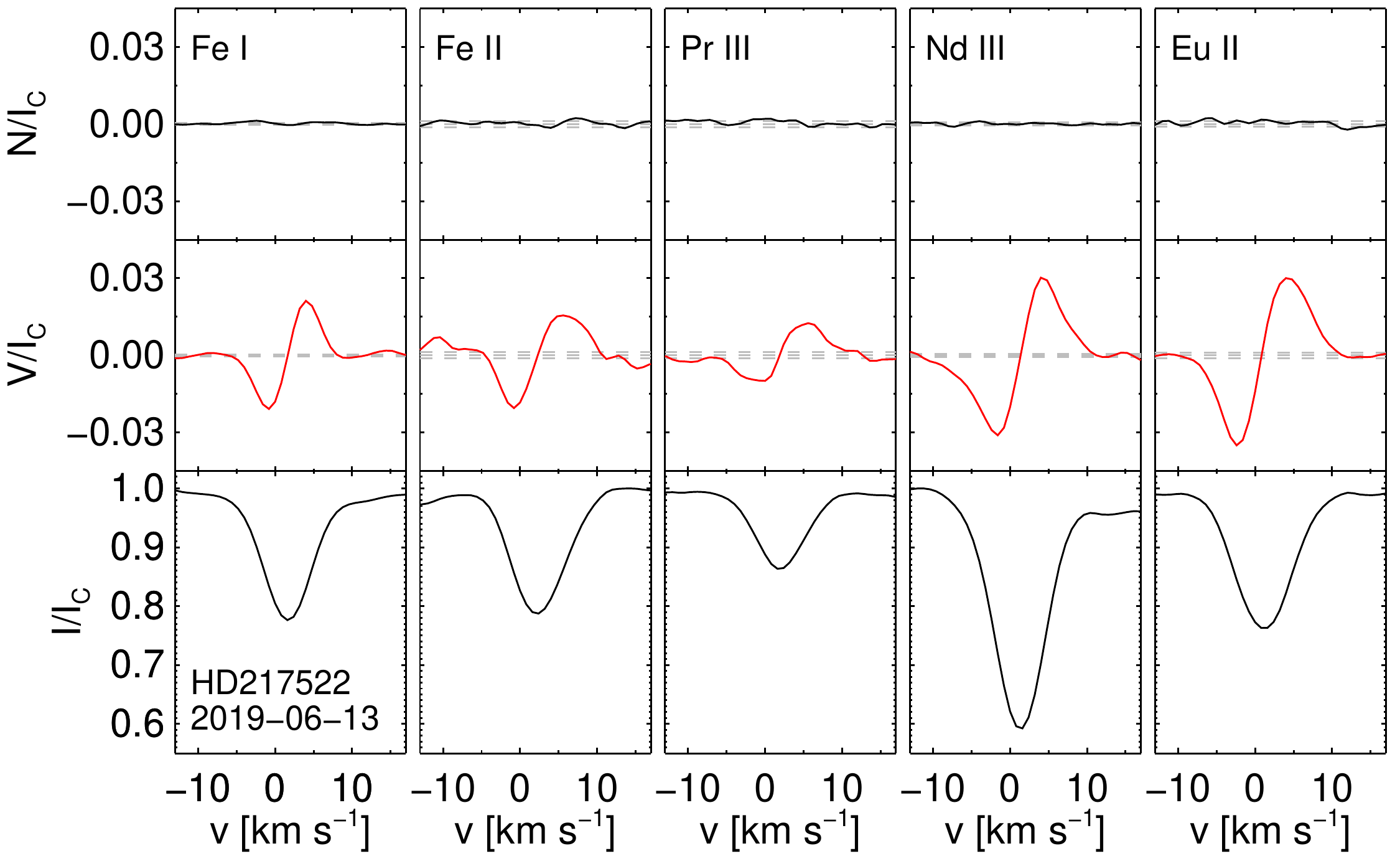}
\includegraphics[width=0.48\textwidth]{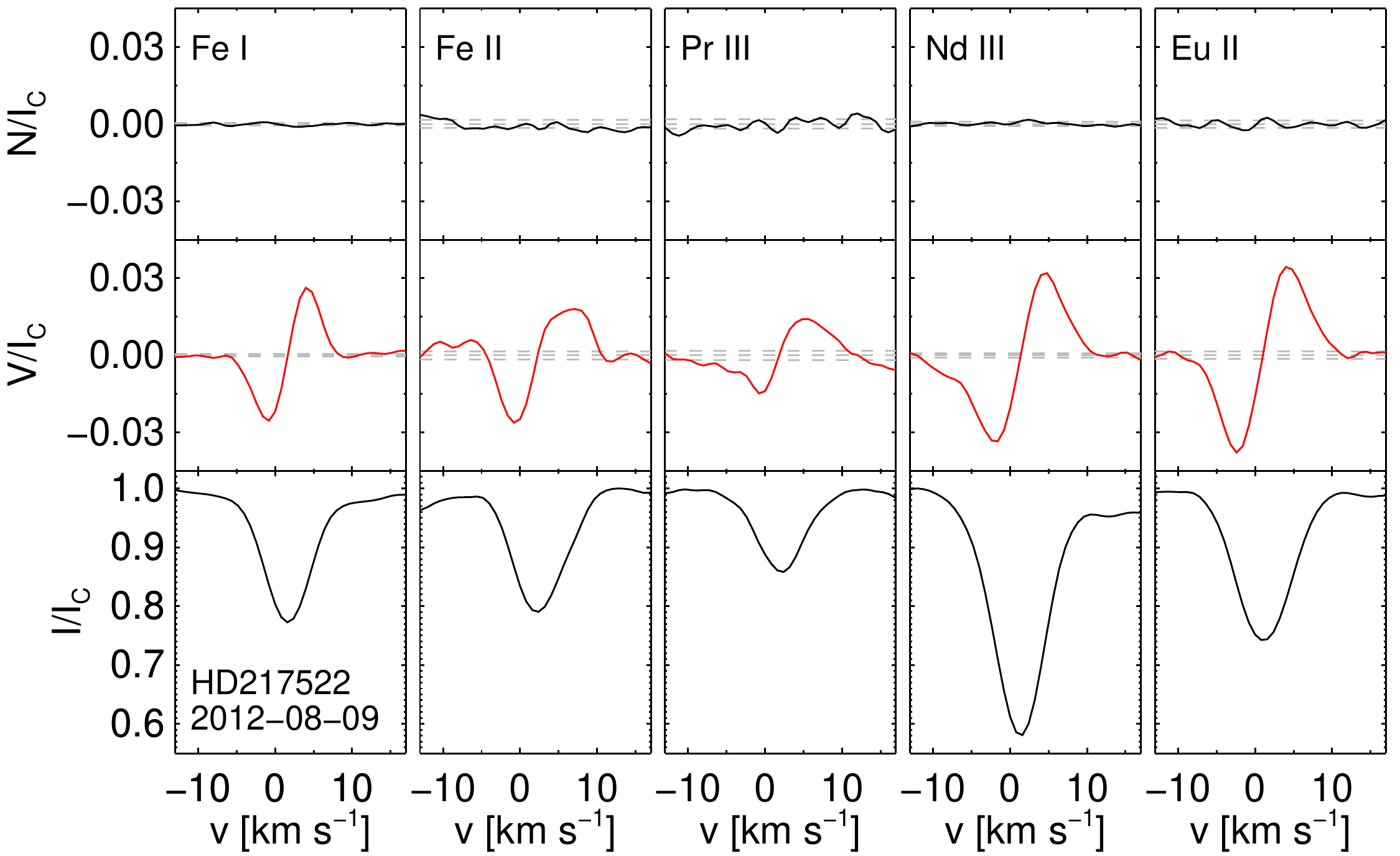}
\caption{
As Fig.~\ref{fig:IVNHD89393}, but for HD\,217522.
}
\label{fig:IVNHD217522}
\end{figure}

Our analysis presented in Tables~\ref{tab:obsall} and \ref{tab:Bzelem} and 
in Figs.~\ref{fig:IVNall} and \ref{fig:IVNHD217522} is based on two 
HARPS\-pol observations, one from 2012 August and the second one from 2019 
June. A comparison of the measurements indicates that the field in the 
measurements using all lines has slightly decreased from  
$\left<B_{\rm z}\right>=-401\pm6$\,G to  
$\left<B_{\rm z}\right>=-323\pm6$\,G. Assuming that the maximum field 
strength was reached in the year 2004, the expected rotation period should 
be at least of the order of a few tens of years. The strongest field 
strength was measured using the \ion{Pr}{iii} line mask, reaching  
$\left<B_{\rm z}\right>=-552\pm59$\,G in the observations acquired in 2012.


\subsection{HD\,70702}

\begin{figure}
\centering
\includegraphics[width=0.48\textwidth]{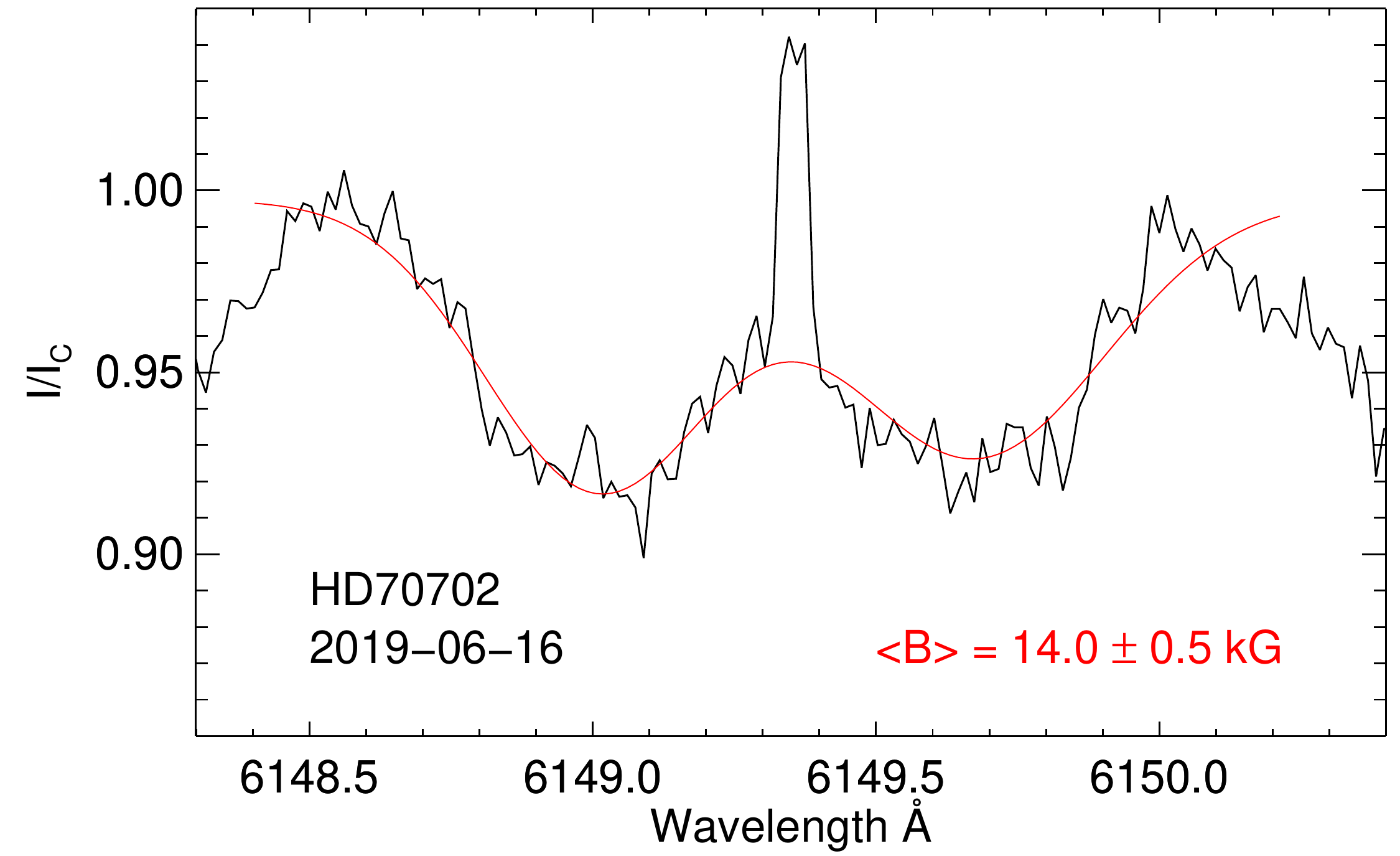}
\includegraphics[width=0.48\textwidth]{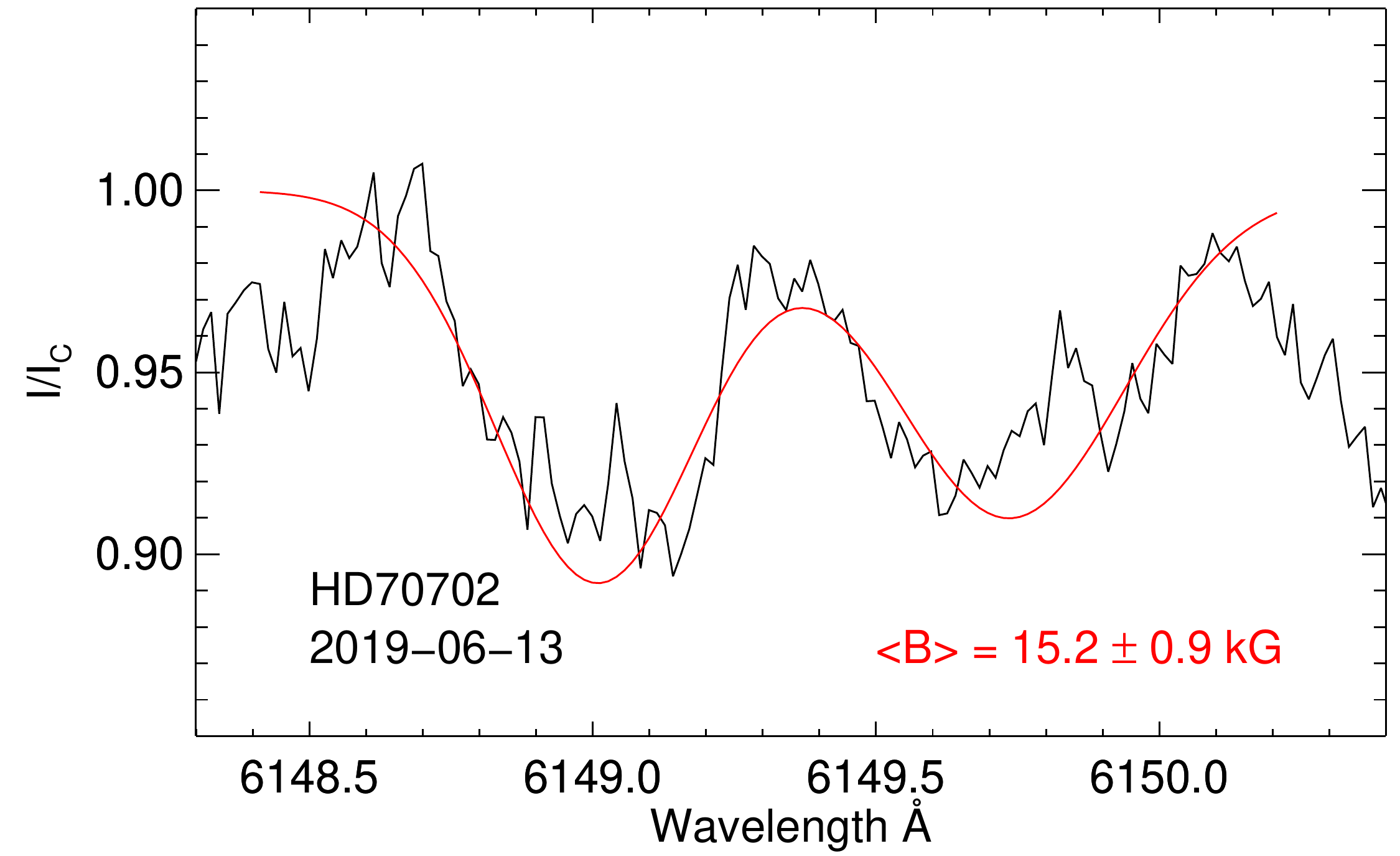}
\caption{
Stokes~$I$ spectra of HD\,70702 in the region of the magnetically split 
\ion{Fe}{ii} line at 6149.258\,\AA{} observed at two different epochs. The 
line profiles are fitted by two Gaussian components represented by the red 
solid lines. The spectrum in the upper panel is affected by a cosmic ray 
appearing between the resolved components (see also the spectra for this 
star in Fig.~\ref{fig:region}).
}
\label{fig:BmodHD70702}
\end{figure}

\begin{figure}
\centering 
\includegraphics[width=0.48\textwidth]{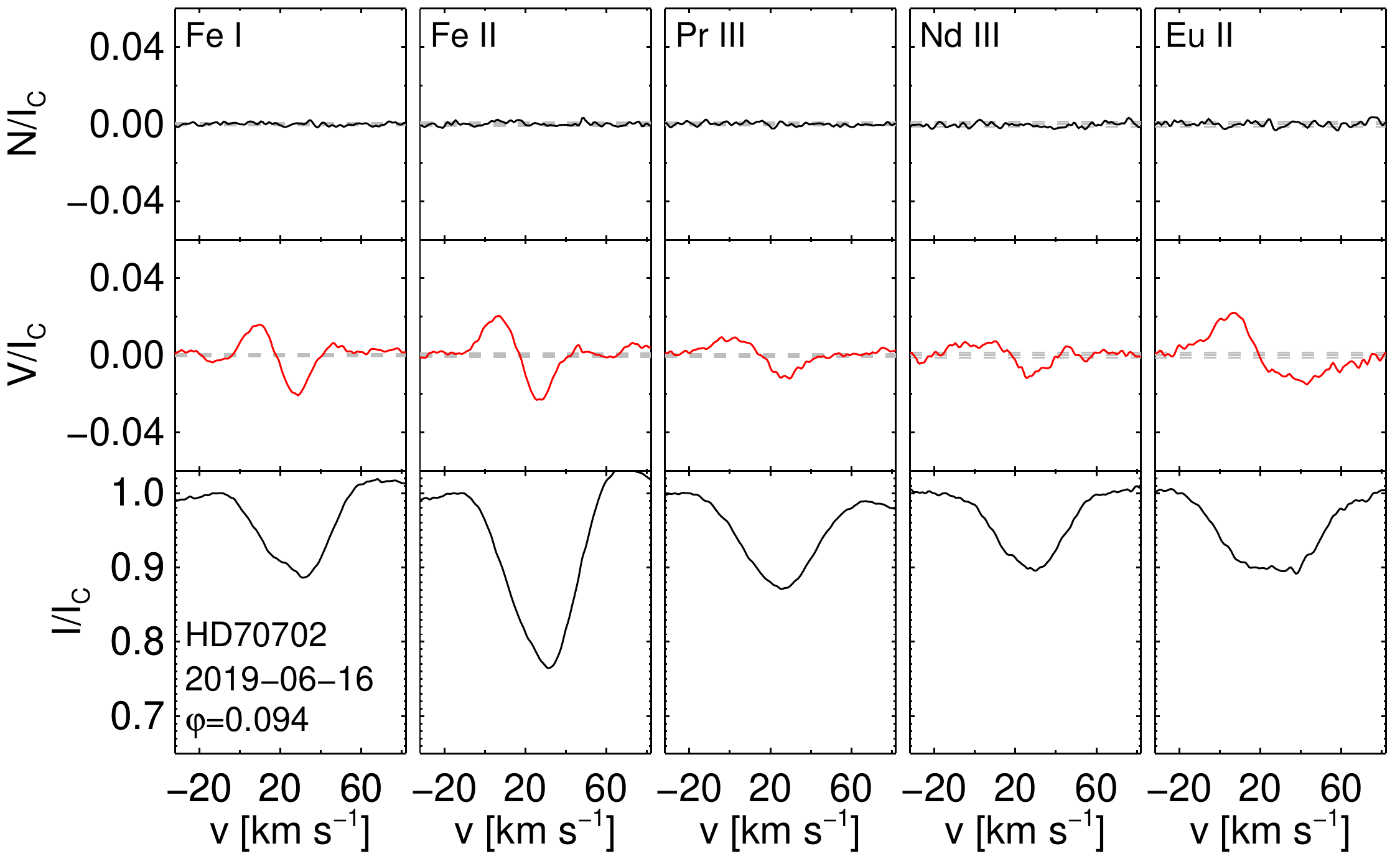}
\includegraphics[width=0.48\textwidth]{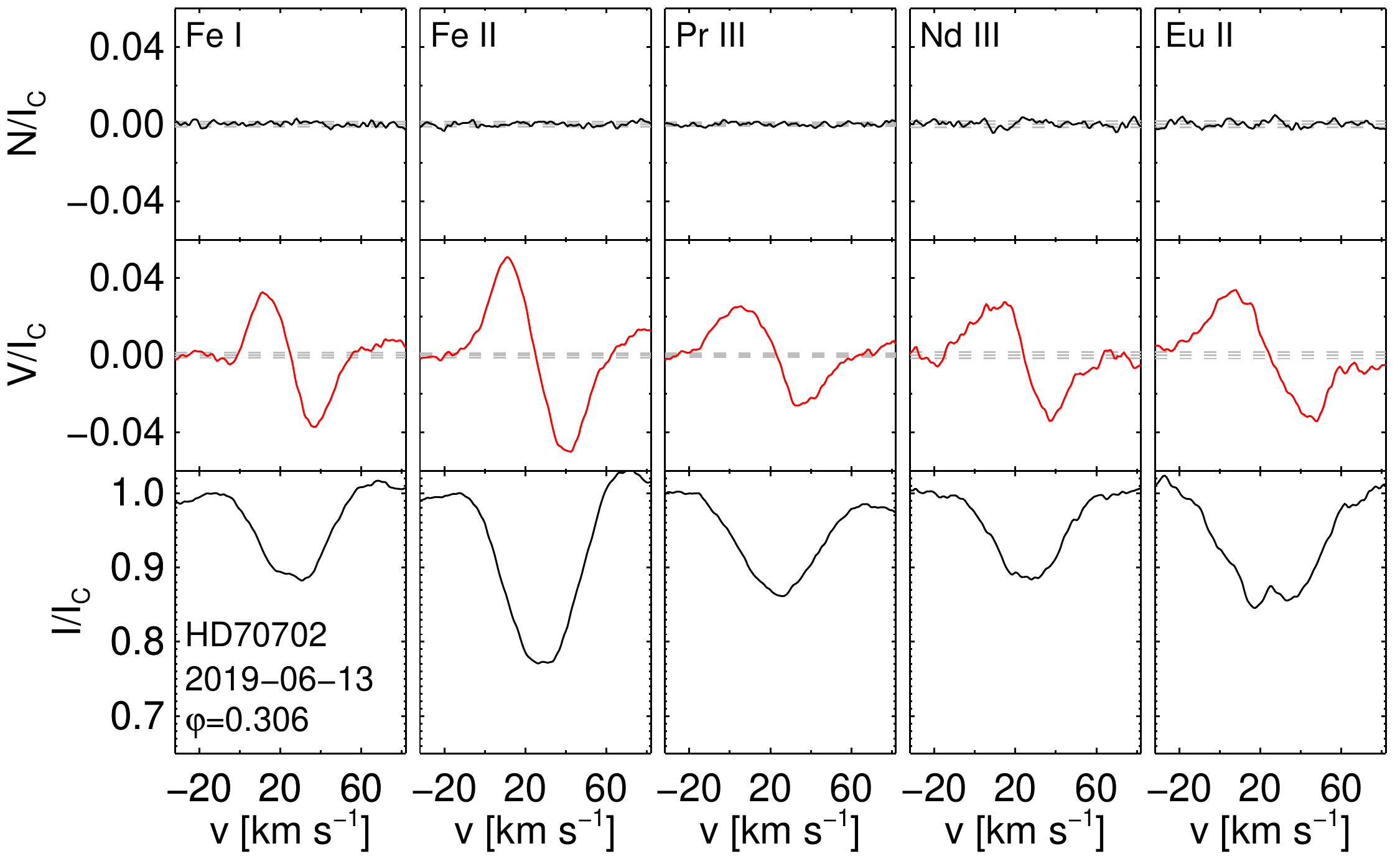}
\caption{
As Fig.~\ref{fig:IVNHD89393}, but for HD\,70702. Since the rotation period is 
known from TESS observations, we also indicate the corresponding 
rotation phases. 
}
\label{fig:IVNHD70702}
\end{figure}

This star is classified as B9\,EuCrSr in 
\citet{Renson} 
and was used in our observations as a standard star. It possesses an extremely 
strong magnetic field with a mean field modulus of the order of 14--15\,kG. 
Because of the strength of this field, numerous lines in the spectra of 
HD\,70702 appear resolved into their magnetically split components
\citep{Elkin}. 
Both of our HARPS spectra are rather noisy and, in addition, the precision of 
the measurements is limited by the distortion of the lines as a result of 
the combination of the Zeeman effect and the rotational Doppler effect. The 
measurements of the mean magnetic field modulus, 
$\left<B\right>=15.2\pm0.9$\,kG (from a spectrum obtained on 
2019 July 13) and $\left<B\right>=14.0\pm0.5$\,kG (from a spectrum obtained 
on 2019 July 16), are presented in Fig.~\ref{fig:BmodHD70702}. They are in 
good agreement with the previously published value of 15\,kG in 
\citet{Elkin}.  
In Fig.~\ref{fig:IVNHD70702}, we present the LSD Stokes~$I$, Stokes~$V$, and 
diagnostic null $N$ profiles calculated for both observing epochs using five 
different masks. The strong variability of HD\,70702 is obvious and is 
clearly reflected in the measurement results presented in 
Tables~\ref{tab:obsall} and \ref{tab:Bzelem}, as well as in the distinct 
changes of the amplitude of the Stokes~$V$ profiles between both epochs. The 
strongest mean longitudinal magnetic field is detected using the 
\ion{Nd}{iii} line mask in the observations from the first epoch, suggesting 
that this element, in comparison to other elements, is concentrated in a 
region of the surface located closer to the magnetic pole.

\begin{figure*}
\centering 
\includegraphics[width=\textwidth]{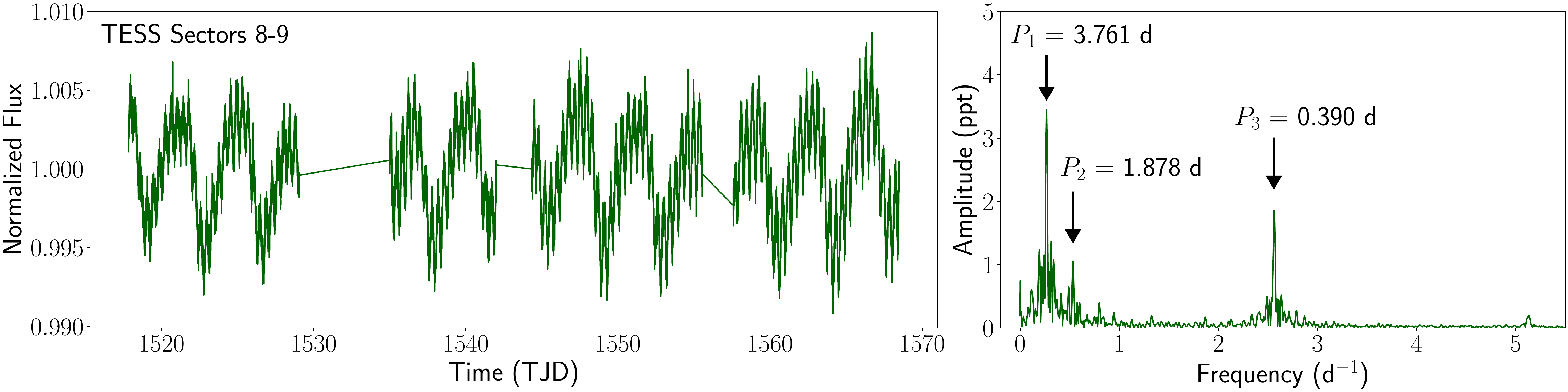}
\caption{
The left panel shows the TESS light curve of HD\,70702 from Sectors 8 and 9. 
The right panel shows the periodogram of this light curve, with the three 
most prominent periods marked. $P_1$ is the rotation period, and $P_3$ is 
likely a pulsation period. Both periods are clearly visible in the light 
curve. $P_2$ represents the first harmonic of $P_1$.
}
\label{fig:PHD70702}
\end{figure*}

In the DFT for this star, we combined the data from Sectors 8 and 9 to find 
three significant periods. Contamination from other stars is likely minimal, 
since the \texttt{CROWDSAP} parameter suggests that only 1.5\% of the flux 
in the optimal aperture does not belong to HD\,70702. We observe a strong 
rotational modulation with $P_{\rm rot} = 3.7601 \pm 0.0007$\,d, as well as 
a short-period modulation -- perhaps due to pulsation -- at $P_{\rm puls} = 
0.39019 \pm 0.00001$\,d. The other significant peak, at 
$P = 1.8782 \pm 0.0002$\,d, has a lower $S/N$ than the others ($\sim5$), but 
still meets our significance criterion. This peak can be identified as a 
harmonic of $P_{\rm rot}$, given that it is exactly twice the frequency (to 
within the uncertainty of the data) of the lower-frequency peak. Using the 
same argument, the small peak visible near $\nu = 5$\,d$^{-1}$ is almost 
certainly a harmonic of $P_{\rm puls}$. The light curve and Fourier 
transform for HD\,70702 are shown in Fig.~\ref{fig:PHD70702}.

Since the TESS data show a clear rotation period, we can calculate the phases
of our HARPS observations. Using as initial epoch $T_{0}$ the value 
TJD\,1564.13520327 (=JD\,2458564.13520327), which corresponds to the time of 
minimum light, we calculate $\varphi=0.306$ for the first observation and 
$\varphi=0.094$ for the second observation. Since this star has a rather 
short rotation period and possesses an extremely strong magnetic field, it 
is an excellent candidate for future spectropolarimetric monitoring to map 
its magnetic field and chemical spots using Zeeman Doppler Imaging.

\section{Discussion and conclusions}
\label{sec:dis}

Among the three sharp-lined stars not previously studied for the presence of 
a magnetic field (HD\,89393, HD\,174779, and HD\,189832), the weakest mean 
longitudinal magnetic field, $\left< B_{\rm z}\right>=-45\pm3$\,G was 
detected in HD\,174779 in the LSD measurements using all lines, i.e.\ based 
on the line mask combining the five individual line masks. The magnetic 
field measurements of HD\,189832 yielded a field strength of 140\,G, whereas 
the field strength for HD\,89393 was 300\,G. Considering the field 
measurements using the five line masks separately, the strongest magnetic 
field was always detected for a line mask corresponding to one of the REEs: 
in the \ion{Eu}{ii} lines for HD\,89393 and in the \ion{Pr}{iii} lines for 
HD\,174779 and HD\,189832. TESS observations of HD\,89393 and HD\,174779 
revealed only the presence of marginal peaks at low $S/N$, indicating that 
the rotation periods of these stars are likely long. HD\,189832 was reported 
to have a $P_{\rm rot}=18.89$\,d from ground-based photometry in
\citet{ManMat}.

The presence of weak magnetic fields in the sharp-lined stars HD\,138633, 
HD\,176196, HD\,203932, and HD\,217522 was already discussed in previous 
studies, but the detections were, in most cases, based on low-resolution 
spectropolatimetric observations. The weakest mean longitudinal magnetic 
field, $\left< B_{\rm z}\right>=21\pm4$\,G, was detected in HD\,203932 using 
all lines for the measurements. Field strengths of the order of 120\,G and 
200\,G were measured for HD\,176196 and HD\,138633, respectively. The mean 
longitudinal magnetic field of HD\,217522 decreased from $-400$\,G to 
$-323$\,G over the seven years from 2012 to 2019. As for the field 
measurements using five line masks separately, similar to the three above 
discussed sharp-lined stars lacking previous magnetic field determinations, 
the strongest magnetic field was always detected in a line list 
corresponding to one of the REEs, in \ion{Pr}{iii} lines for three stars and 
in \ion{Eu}{ii} lines for HD\,203932. These results clearly show that the 
most promising way to detect weak magnetic fields in sharp-lined Ap stars 
is to apply the LSD technique to a line mask belonging to REEs. Our finding 
that stronger fields are detected using REE line masks is probably explained 
by the location of surface REE patches in the vicinity of the magnetic poles.

TESS observations show a short $P_{\rm rot}$ of 6.44\,d for
HD\,203932, which exhibits in the polarimetric spectrum a Zeeman feature
with a typical crossover profile. Longer rotation periods are suggested for 
HD\,89393, HD\,138633, HD\,174779, HD\,176196, and HD\,217522, either from 
TESS observations or spectropolarimetric observations. A very long 
rotation period, on the order of tens of years, is suggested for the strongly 
magnetic roAp star HD\,137949.

Summarising the results of our study, out of seven sharp-lined stars, two 
stars, HD\,174779 and HD\,203932, exhibit a rather weak longitudinal 
magnetic field. We should, however, keep in mind that the dipole strengths for 
these two stars are not known yet, due to the absence of their 
$\left< B_{\rm z}\right>$ phase curves. The fact that the majority of the 
studied stars are slow rotators -- apart from HD\,189832 and HD\,203932, all 
other studied sharp-lined stars have long rotation periods -- is in 
agreement with our expectations that the inclination angles of the rotation 
axes of our sample stars with respect to our line of sight are randomly 
distributed. This is also in agreement with the previous work of 
\citet{Hubrig2007}.
On the other hand, the absence of rotational variability can also be 
expected if the magnetic obliquity $\beta$ with respect to the rotation axis 
is very small 
\citep[e.g.][]{MathysTESS}.
Interestingly, the study by 
\citet{Hubrig2007}
on the evolution of magnetic fields in stars across the upper main sequence  
using a sample of 90 Ap and Bp stars with accurate Hipparcos parallaxes and 
definitely determined
longitudinal magnetic field phase curves revealed that the angle $\beta$ is 
smaller than 20\degr{} in slower rotating stars.

The rotation periods of Ap stars span up to five or six orders of magnitude 
\citep[e.g.][]{MathysSVOS},
but no evidence was previously found for any loss of angular momentum during 
the main-sequence lifetime 
\citep[e.g.][]{Hubrig2007}.
\citet{MathysPASE}
denoted all Ap and Bp stars with rotation periods longer than 50\,d as 
super-slowly rotating Ap (ssrAp) stars and presented accurate periods for 33 
such targets, with the 29\,yr period of HD\,50169 being the longest of them 
\citep{Mathys2019}.
A rotation period of about 35462.5\,d ($\sim97$\,yr) was suggested by 
\citet{gammaEqu}
for the Ap star  $\gamma$\,Equ (=HD\,201601), and of about 188\,yr for 
Przybylski’s star (=HD\,101065) by 
\citet{Hubrig2018}.
It was also suggested that weak-field Ap stars may potentially represent a 
significant fraction of the group of very slowly rotating stars, with rotation 
periods reaching several hundred years 
\citep{MathysSVOS}.
An explanation for the extremely slow rotation of a fraction of Ap stars is 
currently missing, although a first theoretical attempt to understand this 
phenomenon was recently presented by 
\citet{Kitchatinov},
who suggested a possible scenario in terms of a longitudinal drift of the 
unstable disturbances of a kink-type (Tayler) instability of the stellar 
internal magnetic field.

Our studied sample of sharp-lined Ap stars is rather small and does not 
allow us to decide whether a critical value for the stability of a 
large-scale magnetic field indeed exists, or if previous observations are 
incomplete, insofar as they are missing a sizeable population of chemically 
peculiar sharp-lined stars without any magnetic or Doppler line broadening. 
Therefore, additional systematic, multi-epoch $\bz$ determinations for the 
known Ap and Bp stars with sharp unresolved spectral lines are needed to 
characterise the distribution of the slowly rotating Ap stars, both in terms 
of their magnetic field strengths and rotation periods, in order to provide 
essential clues for the theoretical understanding of the formation and 
evolution of these stars.


\section*{Acknowledgements}

We thank the anonymous referee for their helpful comments.
We also thank Gautier Mathys for the discussion on several stars in our 
sample.
Based on observations made with ESO Telescopes at the La Silla Paranal 
Observatory under programme IDs 68.D-0445, 072.D-0138, 089.D-0383, and 
0103.C-0240. 
This paper includes data collected by the TESS mission. Funding for the
TESS mission is provided by the NASA Science Mission Directorate. Resources
supporting this work were provided by the NASA High-End Computing (HEC) Program
through the NASA Advanced Supercomputing (NAS) Division at Ames Research Center
to produce the SPOC data products. 
This work has made use of the VALD, operated at Uppsala University, the
Institute of Astronomy RAS in Moscow, and the University of Vienna.


\section*{Data Availability}

The data obtained with ESO facilities are available in the ESO Archive at
http://archive.eso.org/ and can be found with the instrument and
object name.

TESS light curves are publicly available through the Mikulski
Archive for Space Telescopes 
(https://mast.stsci.edu/portal/Mashup/Clients/Mast/Portal.html).



\label{lastpage}

\end{document}